\documentclass[11pt,onecolumn,draftcls]{IEEEtran}
 \usepackage{setspace}
\usepackage{graphicx}
\usepackage{amsmath,amssymb,epsfig}
\usepackage{varioref}
\usepackage{subfigure}
\usepackage[sort&compress,square,numbers]{natbib}
\usepackage{blkarray}
\usepackage[bookmarks]{hyperref}
\usepackage{arydshln}
\usepackage{enumerate}
\usepackage{color}
\usepackage{epsfig}
\newenvironment{typew}{\sffamily}{}
\newcommand{\btw}{\begin{typew}}
\newcommand{\etw}{\end{typew}}

\newtheorem{theorem}{Theorem}
\newtheorem{proposition}[theorem]{Proposition}
\newtheorem{lemma}[theorem]{Lemma}
\newtheorem{corollary}[theorem]{Corollary}

\newtheorem{claim}[theorem]{Claim}

\newtheorem{definition}{Definition}

\newcommand{\beq}{\begin{equation}}
\newcommand{\eeq}{\end{equation}}
\newcommand{\bea}{\begin{array}}
\newcommand{\ena}{\end{array}}
\newcommand{\bds}{\begin {itemize}}
\newcommand{\eds}{\end {itemize}}
\newcommand{\bdf}{\begin{definition}}
\newcommand{\blm}{\begin{lemma}}
\newcommand{\edf}{\end{definition}}
\newcommand{\elm}{\end{lemma}}
\newcommand{\bthm}{\begin{theorem}}
\newcommand{\ethm}{\end{theorem}}
\newcommand{\bprp}{\begin{prop}}
\newcommand{\eprp}{\end{prop}}
\newcommand{\bcl}{\begin{claim}}
\newcommand{\ecl}{\end{claim}}
\newcommand{\bcr}{\begin{coro}}
\newcommand{\ecr}{\end{coro}}
\newcommand{\bquest}{\begin{question}}
\newcommand{\equest}{\end{question}}

\def\bm#1{\mbox{\boldmath $#1$}}

\newcommand{\bSigma}{\mbox{$\bm \Sigma$}}

\newcommand{\yvec}{{\bf{y}}}

\newcommand{\wvec}{{\bf{w}}}
\newcommand{\xvec}{{\bf{x}}}

\newcommand{\rvec}{{\bf{r}}}

\newcommand{\vvec}{{\bf{v}}}

\newcommand{\Amat}{{\bf{A}}}
\newcommand{\Bmat}{{\bf{B}}}

\newcommand{\Emat}{{\bf{E}}}

\newcommand{\Gmat}{{\bf{G}}}
\newcommand{\Hmat}{{\bf{H}}}

\newcommand{\Imat}{{\bf{I}}}

\newcommand{\Tmat}{{\bf{T}}}

\newcommand{\Rmat}{{\bf{R}}}
\newcommand{\Umat}{{\bf{U}}}
\newcommand{\Vmat}{{\bf{V}}}
\newcommand{\Wmat}{{\bf{W}}}

\newcommand{\comp}{{\mathbb{C}}}
\newcommand{\nat}{{\mathbb{N}}}

\newcommand{\0}{{\mathbf {0}}}

\newcommand{\define}{\stackrel{\triangle}{=}}

\def\bLambda{{\mbox{\boldmath $\Lambda$}}}

\def\tvec{{\mbox{\boldmath $t$}}}

\newcommand{\be}{\begin{equation}}
\newcommand{\ee}{\end{equation}}
\newcommand{\beqna}{\begin{eqnarray}}
\newcommand{\eeqna}{\end{eqnarray}}

\usepackage{algorithm}
\usepackage{algorithmic}

\title{The One-Bit Null Space Learning\\ Algorithm  and its Convergence     }

\author{Yair Noam and Andrea J. Goldsmith, \IEEEmembership{Fellow, IEEE}\thanks{This work is supported by the ONR under grant N000140910072P00006,
the AFOSR under grant FA9550-08-1-0480, and the DTRA under grant HDTRA1-08-1-0010.

 Yair Noam is with the Faculty of Engineering, Bar-Ilan University,
Ramat-Gan, 52900, Israel.

 Andrea J. Goldsmith is with the Dept. of Electrical Engineering, Stanford University, Stanford CA, 940305.}}

\begin{document}

\maketitle
\date{}

\begin{abstract}
 This paper proposes a new algorithm for MIMO cognitive radio Secondary Users (SU) to learn the null space of the interference channel to the Primary User (PU)  without  burdening the PU with any  knowledge  or explicit cooperation   with the  SU.
 The knowledge of this null space enables the SU to transmit in the same band   simultaneously  with the PU by  utilizing separate spatial dimensions  than the PU. Specifically, the SU transmits in the null space of the interference channel to the PU.  We present a new algorithm, called the One-Bit Null Space Learning Algorithm (OBNSLA), in which   the   SU   learns the PU's null space by  observing a binary function that indicates whether  the interference it inflicts on the PU has increased or decreased in comparison to the SU's previous  transmitted signal. This function is obtained  by listening to the PU transmitted signal or control channel and  extracting   information from it about whether the   PU's Signal to Interference plus Noise power Ratio (SINR) has increased or decreased.

In addition to introducing the OBNSLA, this paper provides a thorough convergence analysis of this algorithm. The OBNSLA is shown to have a linear convergence rate and an asymptotic quadratic convergence rate. Finally,  we    derive bounds on the interference that the SU inflicts on the PU as a function of a parameter  determined  by the SU. This  lets  the SU     control   the  maximum level  of interference, which enables it to protect the PU completely blindly with minimum complexity. The asymptotic analysis and the derived bounds  also apply to the recently proposed Blind Null Space Learning Algorithm.
  \end{abstract}


\section{Introduction}

\begin{figure}
\centering
\epsfig{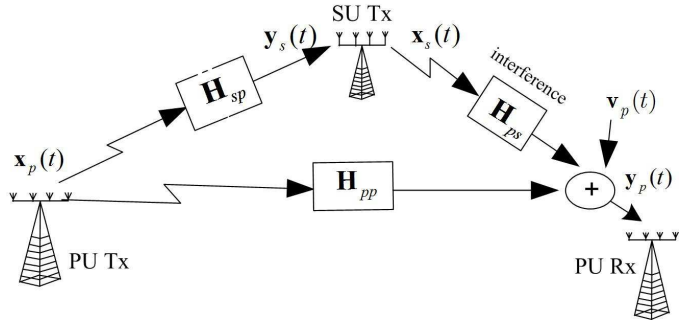}
\caption{Our  cognitive radio scheme.    $\Hmat _{ps}$ is unknown to the secondary transmitter and $\vvec_{p}$ (t) is a stationary noise   (which may include  stationary interference).   The  interference from the SU,
 ${\bf H}_{ps}\xvec_s(t)$,
 is  treated  as noise; i.e., there is no interference cancellation.}
\label{Figure:PassivePrimarysystem} \end{figure}

Multiple Input Multiple Output (MIMO) communication opens new directions  and possibilities for  Cognitive Radio (CR) networks
\citep{ZhangExploiting2008,
ScutariCognitive2008,ScutariMIMO2010,GoldsmithBraking2009,HaykinCognitive2005,
zhang2010cognitive,zhang2010spectrum}. In particular, in underlay CR networks, MIMO technology enables the SU to transmit a significant amount of power   simultaneously in the same band as the Primary User (PU) without interfering with it, if the SU utilizes separate spatial dimensions than the PU. This spatial separation requires that the interference channel from the SU to the PU be known to the SU. Thus, acquiring  this knowledge, or   operating without it, is a major topic of active research in CR
\citep{HuangDecentralized2011,
ZhangRobust2009,zhang2010active,
zhang2010cognitive,ZhangOptimal2011,
ChenInterference2011,YiNullspace,Noam2012Blind,Nehorai2006} and in other fields \citep{mudumbai2005scalable}.
    We consider MIMO primary and secondary systems defined as follows: we assume a flat-fading   MIMO channel with one PU and one SU, as depicted in Fig. \ref{Figure:PassivePrimarysystem}.   Let ${\bf H}_{ps}$   be  the  channel matrix between  the SU's transmitter   and the PU's  receiver, hereafter   referred to  as the SU-Tx and PU-Rx, respectively.  In the underlay CR paradigm,   SUs   are    constrained   not to inflict ``harmful"'' interference on the PU-Rx. This can be achieved if the SU restricts its signal to lie within the null space of \(\Hmat_{ps}\);  however, this is only possible if the  SU   knows $\Hmat_{sp}$.
The optimal power allocation in  the  case where the SU  knows the  matrix
 $\Hmat_{ps}$
 in addition to its own Channel State Information (CSI) was derived by Zhang and Liang
 \citep{ZhangExploiting2008}.
 For the case of multiple SUs,  Scutari at al.
 \citep{ScutariMIMO2010}
formulated a  competitive game between the secondary users. Assuming that the interference matrix to the PU is known by each SU, they derived conditions  for the existence and uniqueness of a Nash Equilibrium point to the game. Zhang et al.
 \citep{ZhangRobust2009}
 were the first to take into consideration the fact that the  interference matrix
$\Hmat_{ps}$
may not be  perfectly known  (but is partially known) to the SU. They  proposed robust beamforming  to assure compliance with the interference constraint of the PU while maximizing the SU's
 throughput. Another   work  on the case of an unknown interference channel with known probability distribution is due to
Zhang and So \citep{ZhangOptimal2011}, who optimized the SU's throughput under a constraint on the  maximum probability that the interference to the PU is above a threshold.

The underlay concept  of CR in general, and MIMO CR in particular,  is  that the SU must be able to mitigate the interference to the PU  blindly without any cooperation. 
Zhang \citep{zhang2010cognitive} was the first to propose a    blind solution  where the MIMO SU mitigates interference to the PU by null space learning. This work was followed by   Yi
 \citep{YiNullspace}, Chen et al. \citep{ChenInterference2011}, and Gao et al. \citep{Nehorai2006}. All these works exploit  channel reciprocity: specifically, where the SU listens to the PU's transmitted signal and estimates the null space    from the signal's  second order statistics. Since  these works require channel reciprocity, they are restricted to PUs that use Time Division Duplexing (TDD).

 Unless there is channel reciprocity, obtaining $\Hmat_{ps}$ by the SU requires cooperation with the PU in the   estimation  phase; e.g. where the SU transmits  a training  sequence, from which the PU  estimates $\Hmat_{ps}$ and feeds it back to the  SU.  Cooperation of this nature     increases the system complexity overhead, since it requires a handshake between both systems and, in addition,   the PU needs to be synchronized with   the SU's training sequence. Zhang \citep{zhang2010active} was the first to propose an interference mitigation mechanism
in which  a single antenna   SU obtains the path-loss of  the interference channel to the PU under the condition that the SU can    extract
the  PU's Signal to Interference plus Noise Ratio  (SINR) by listening to its transmitted signal or control channel. By transmitting an interfering signal, and measuring the effect of this signal on the PU SINR, the SU obtains the path-loss of the interference channel. This
enables the SU to set its power low enough to maintain its interference
below some predefined level. However, it does not enable the
SU  to exploit
other spatial degrees of freedom than those used by the PU. In \citep{Noam2012Blind}, we proposed the Blind Null Space Learning Algorithm (BNSLA), which  enables a MIMO underlay CR to learn the null space of \(\Hmat_{ps}\)  by observing some unknown monotone
continuous function of the PU's  SINR. For example, if the PU is using continuous power control, the PU's signal power is a monotone function of its SINR.
 During this learning, the PU does not cooperate at all with the  SU  and operates  as though there were no other systems in the medium (the way current PUs operate today).   

 This paper makes two contributions. The first contribution is a  new algorithm, called the One-Bit Null Space Learning Algorithm (OBNSLA), which requires much less information than the BNSLA; namely,   the SU   can infer whether the interference it inflicts on the PU has increased or decreased compared to a  previous time interval with a one-bit function. In other words,    in the OBNSLA the SU measures a one-bit function of the PU's SINR, rather than a continuous-valued function as in the BNSLA. Using this single bit of information, the    SU    learns $\Hmat_{ps}$'s  null space  by iteratively modifying the spatial orientation of its   transmitted signal and measuring  the effect of this modification on the PU's SINR. The second contribution of the paper is to provide a thorough convergence analysis of the OBNSLA. We show that the algorithm converges linearly and has an asymptotically quadratic convergence rate. In addition, we derive upper bounds   on the interference that the SU inflicts on the PU; these results enable the SU to control the interference to the PU without any cooperation on its part. Furthermore, all the bounds and the convergence results apply  equally to the BNSLA.


\section{  The One-Bit Null Space Learning Problem} \label{Section:ProblemFormulation}

Consider a flat fading MIMO interference channel  with  a single PU  and a single  SU  without interference cancellation; i.e., each system treats the other system's signal as noise. The PU's received signal is
\beq \label{PU Observed Signal}
\yvec_p(t) = \Hmat_{pp} \xvec_p(t) + \Hmat_{ps}\xvec_{s}(t) +\vvec_p(t),\; t\in {\mathbb N}
\eeq
  where \(\xvec_{p}\), \(\xvec_{s}\) is the PU's and SU's transmitted signal, respectively,  \(\Hmat_{pp}\) is the PU's direct channel; \(\Hmat_{ps}\) is the interference channel between the PU Rx and the SU Tx, and $\vvec_p(t)$ is a zero mean stationary noise. In the underlay CR paradigm,   the SU   is    constrained not to exceed a maximum interference level at  the PU Rx; i.e.,
\beq
\label{InterferenceConstraint}
\|\Hmat_{ps}\xvec_{s}(t)\|^{2}\leq \eta_{\rm max}, \eeq
where \(\eta_{{ \rm max} }>0\) is the maximum interference constraint.
In this paper, all vectors are column vectors. Let $\Amat$ be an $l\times m$ complex matrix; then, its null space is  defined as  $ {\cal N}(\Amat)=\{\yvec\in \mathbb C^{m}:\Amat \yvec=\0 \}$ where $\0=[0,...,0]^{\rm T}\in \comp^{l}$.

\begin{figure}
\centering
{ \epsfig{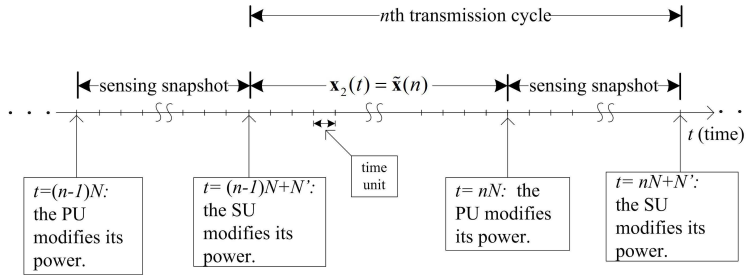}}
\caption{
The time indexing  used in this paper. $t$  indexes  the basic time unit (pulse time) where \(N\) time units constitute a TC that is indexed by $n$.
Furthermore,  \(K\) transmission cycles constitute a learning phase (not shown  in this figure).}
\label{Figure:index}
\end{figure}

  Since our focus is on constraining the interference caused by the SU to the PU, we only consider the term \(\Hmat_{ps}\xvec_{s}(t)\) in \eqref{PU Observed Signal}.  Hence, $\Hmat_{ps}$ and \(\xvec_{s}\) will be denoted by  $\Hmat$ and \(\xvec\), respectively.
 We also define the   Hermitian matrix \(\Gmat\) as \beq\label{DefineG} \Gmat= \Hmat^* \Hmat \eeq  The time line $\mathbb N$ is   divided into  $N$-length intervals, each  referred to as a transmission cycle (TC), as depicted in Figure \ref{Figure:index}.  For each TC, the SU's signal  is constant;  i.e., \beq\label{SecondarySignal}
 \begin{array}{lll}\xvec_{s}((n-1)N+N') =\xvec_{s} ((n-1)N+1) =
\cdots=\xvec_s(Nn+N'-1)\triangleq\tilde
\xvec(n),\end{array}\eeq where the time interval  $nN<t\leq nN+N'-1$  is the snapshot in which the SU measures  a  sequence \(q(n),\) where each \(q(n)\) is a function of the interference that the SU inflicts on the PU.  We assume that the SU can extract one-bit of information  from the sequence  \(q(n),\) which indicates whether the interference it inflicts on the PU, at the \(n\)th TC,  has increased or decreased with respect to the previous TCs.   This assumption is described in the following.
\paragraph*{Observation Constraint (OC) on the function \(q(n)\)}{ Let \(q(n), ~n=1,2,...\) be a sequence observed by the SU where \(n\) is the index of a TC, and let \( \Hmat\tilde \xvec(n)\) be the interference that the SU inflicts on the PU at the \(n\)th TC. Then, \(q(n)\) is a function of   the interference that the SU inflicts on the PU as follows:   There exists  some integer \(M\geq 1\), such that from    $q(n-m),...,q(n)$, the SU can extract the following function   \beq\label{Define H tilde}\tilde h(\tilde \xvec(n),\tilde\xvec(n-m) )=\left\{\begin{array}{l}1, \text{ if } \Vert \Hmat\tilde \xvec(n)\Vert\geq \Vert \Hmat\tilde \xvec(n-m)\Vert\\ -1, \text{ otherwise}\end{array}\right.\eeq for every $m\leq M$. }

\begin{figure}
\centering
{ \epsfig{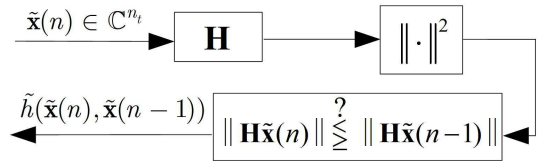}}
\caption{Block Diagram of the One-bit Blind Null Space Learning Problem. The SU's objective is to learn the null space of $\Hmat$ by inserting a series of $\{\tilde\xvec(n)\}_{n\in {\mathbb N}}$  and measuring     \(\tilde h(\tilde \xvec(n),\tilde\xvec(n-1) )\) as output.  In practice, the SU  does not measure \(\tilde h(\tilde \xvec(n),\tilde\xvec(n-1) )\) directly, but  rather it measures a sequence  \(q(n)\), for each \(\tilde \xvec(n)\), and from \(q(n),q(n-1)\), the SU extracts   \(\tilde h(\tilde \xvec(n),\tilde\xvec(n-1)). \)     }
\label{Figure:BlindLearning}
\end{figure}

  The SU's objective is to   learn ${\cal N}(\Hmat)$ from   $ \{\tilde\xvec(n),q (n)\}_{n\in {\mathbb N}}$. This problem, referred to as the One-bit Blind Null Space Learning (OBNSL) problem, is  illustrated in Figure \ref{Figure:BlindLearning} for \(M=1\).
The OBNSL problem is  similar  to the Blind Null Space Learning (BNSL) problem  \cite{Noam2012Blind} except for one important difference. In the latter, the SU  observes  a continuous-valued  function of the PU's SINR  whereas in the OBNSL problem, it observes a  one-bit function. In both problems, the SU  obtains \(q(n)\)  by  measuring the  PU's transmit energy, or any other parameter that indicates the PU's SINR (see Sec. II-B in \citep[]{Noam2012Blind}  for examples).   However, in the  OBNSL problem, the SU is more flexible since it can obtain \(q(n)\) from, for example, incremental power control\footnote{This is power control that is carried out using one-bit command which indicates weather to increase or decrease the power by a certain amount.} or other   quantized functions of the PU's SINR such as modulation size.
  Another way for the SU to extract information about the interference to the
PU is by decoding the PU's control signal to obtain
parameters such as channel quality indicator feedback or ACK/NAK feedback \citep{HuangDecentralized2011}.
 From a system point of view, the OC
 means that between $m$  consecutive transmission  cycles,  the   PU's SINR is mostly  affected by   variations in the    SU's signal. Note that the  OC is less restrictive for smaller values of \(m\). The TC length is the minimum time it takes the SU to modify its learning signal. This length must be equal or greater to the
PU's interference adaptation interval for the learning to be accurate. In addition, variations in other sources of interference and
 in the PU's direct channel should occur on a much slower timescale than the TC length or else the learning may not converge.  It is important to stress that the latter constraint  applies only  to the TC time and not  to the entire time it takes the SU to learn the null space of \(\Hmat\). This is because the OBNSL problem is based only on the variation in the interference with respect to the previous TC,  and if \(M>1\) it is with respect to the variations in the  \(M\)th previous TCs. It is therefore possible that the environment and the PU's direct channel vary faster than \(\Hmat\) as long as these variations are slow with respect to the TC.      In Sec. \ref{Section:Simulation} we study the effect of  a time-varying environment on  the proposed learning scheme, via simulation, and  show that it is possible to learn the null space even when the PU direct channel varies faster then \(\Hmat\). Note that in the case where   \(q(n)\) is not extracted from the PU's SINR,
the PU's path-loss may not affect the OC. Consider  the case
where  the  PU constantly  measures  the  interference  power, or the interference spatial covariance matrix  at the Rx and  feeds it back to its Tx.
For example, such a mechanism is necessary if the PU has full CSI at its Tx. In this case, if the  SU can decode the PU's  control signal,  it can  extract \(q(n)\) from it without being  affected by variations in the PU direct channel; i.e., \(\Hmat_{pp}(t)\).

    The learning process unfolds  as follows.  In the first   TC ($n=1$),  the SU transmits $\tilde \xvec(1)$,  and measures $q(1)$.  In the next TC, the SU transmits   $\tilde\xvec(2)$   and  measures $q(2)$ from which it extracts $\tilde h(\tilde \xvec(1),\tilde \xvec(2))$. This process is repeated until the null space is approximated. Note that while $\tilde h(\tilde \xvec(1),\tilde \xvec(2))$ requires two TCs, $\tilde h(\tilde \xvec(n-1),\tilde \xvec(n))$ for $n>2$ requires a single TC.
Note that the OC does not provide  an explicit relation between $\tilde h(\tilde\xvec(n),\tilde\xvec(n-m))$ and \(q(n)\). This is because  the way  $\tilde h(\tilde\xvec(n),\tilde\xvec(n-m))$ is extracted from \(q(n)\) depends on the PU''s communication protocol. For example, if the PU is using  incremental power control,  and the SU observes these power control commands; i.e.,  $q(n) $ is equal to the PU's power command at the $n$th TC, then $h (\tilde\xvec (n),\tilde \xvec (n-1)) $ will be equal to $q(n) $, and $M$  will be equal to one. On the other hand, if the PU is using a continuous power control and the \(q(n)\) that the SU observes is some monotone function of the PU's power\footnote{See \citep[Sec. II-B]{Noam2012Blind}  for examples and conditions under which this is possible.}, then $\tilde h (\tilde\xvec (n),\tilde\xvec (n-m)) $ will be the difference between $q(n) $ and $q(n-m) $. The value of $M$ in this case will be the number of TCs in which the PU's direct channel and the interference from the rest of the environment remains constant.

\section{ The One-Bit Blind Null Space Learning  Algorithm (OBNSLA)}
\label{Section:Mathematical Descreaption of the Blind Jacobi Technique}

We now present the OBNSLA by which the  SU   approximates
  ${\cal N}(\Hmat)$ from   $ \{\tilde\xvec (n),q_{} (n)\}_{n=1}^{T}$ under the OC, where the approximation error can be made arbitrarily small for sufficiently large  $T$.
Once the SU learns \({\cal N}(\Hmat)\), it can optimize its transmitted signal, regardless of   the optimization  criterion, under the constraint that its  signal   lies in ${\cal N}(\Hmat)$.
Let $\Umat \bSigma \Vmat^*$ be  $\Hmat$'s   Singular Value Decomposition (SVD), where $\Vmat$  and $\Umat$ are  $n_{t}\times n_{t}$ and $n_{r} \times n_{r}$ unitary matrices, respectively, and assume that \(n_{t}>n_{r}\). The matrix   $\bSigma$ is an $n_{r}\times n_{t}$ diagonal matrix with real nonnegative  diagonal entries $\sigma_{1} ,...,\sigma_ {d}$    arranged as  $\sigma_{1}\geq\ \sigma_ {2},\geq \cdots\geq\sigma_ {d}>0$. We assume without loss of generality that \(n_{r}=d(={\rm Rank}(\Hmat))\). In this case \(
{\mathcal N}(\Hmat)={\rm span} (\vvec_{n_{r}+1} ,...,\vvec_ {n_{t}}),
\)
where $\vvec_{i}$ denotes $\Vmat$'s $i$th column. From the SU's point of view, it is sufficient to learn ${\cal N}(\Gmat)$ (recall,  \(\Gmat=\Hmat^{*}\Hmat\)), which is equal to ${\cal N}(\Hmat)$ since  \beq\label{EVD of G}\Gmat= \Vmat \bLambda \Vmat^*,\eeq where $\bLambda=\bSigma ^{\rm T}\bSigma$.
The decomposition in \eqref{EVD of G} is known as the Eigenvalue Decomposition (EVD) of \(\Gmat\). In order to obtain \({\cal N}(\Hmat)\) it is sufficient to obtain \(\Gmat\)'s EVD. However, in the OBNSL problem,  \(\Gmat\) is not observed, so the SU needs to obtain the EVD using only one-bit information.

  To illustrate that \({\cal N}(\Hmat)\)
can be obtained via only one bit  we consider
 a simple example in which \(\Hmat=\alpha[\sqrt3,-1]^{\rm T}\)
 where \(\alpha>0\). In this case \beq\Gmat=\frac{\alpha^{2}}{4}\begin{bmatrix}3 & -\sqrt3 \\
-\sqrt 3 & 1 \\
\end{bmatrix}, \bLambda=\begin{bmatrix}\alpha^{2} & 0 \\
0 & 0 \\
\end{bmatrix},~\Vmat=\frac{1}{2}\begin{bmatrix}-\sqrt3 & 1 \\
1 & \sqrt3 \\
\end{bmatrix}\eeq Note that the null space is spanned by \(\rvec(-\pi/6)\)
where \(\rvec(\theta)=[\sin(\theta),-\cos(\theta)]\). Thus, the null space
can be obtained by minimizing \(\rvec(\theta)^{\rm T}\Gmat\rvec(\theta)\)
over \(\theta\in[-\pi,\pi]\).
 The latter is true due to the fact that \(\rvec(-\pi/6)^{\rm T}\Gmat\rvec(-\pi/6)=0\)
 is the global minimum of the function \(\rvec(\theta)^{\rm T}\Gmat\rvec(\theta)\).  Because \(\rvec(\theta)^{\rm T}\Gmat\rvec(\theta)= \left(\sqrt{3} \sin (\theta )+\cos (\theta )\right)^2/4\) is a sinusoidal function with a period of
\(\pi\), it is possible to search for the null space by transmitting \(\rvec(\theta)\)
for different values
of \(\theta\) and receiving the
one
bit information given in \eqref{Define H tilde}, with linear complexity; i.e., a
complexity that grows with \(1/\epsilon,\) where \(\epsilon\) is the desired accuracy. The extension of the above idea
to practical MIMO channels, which  are complex  and might have more than two antennas, poses some challenges. The first challenge is that because each search point is obtained
via a TC, it is highly desirable to reduce the search complexity. The second problem
is that even for the same dimensions, i.e., a one dimensional null space in \(\comp^2\),
the null space cannot be parameterized by a single parameter \(\theta\) via \(\rvec(\theta)\) but must be parameterized by two parameters, e.g. \((\theta,\phi)\) via \(\rvec(\theta,\phi)=[\cos(\theta),e^{-i\phi}\sin(\theta )]\). Thus, it is necessary to perform an efficient two dimensional search based
on the one bit of information in \eqref{Define H tilde}. This problem is even more
complicated when the dimension of the null space is greater than one. In this
section we address these issues and present the OBNSLA.  In the case of \(n_t>2,\) to avoid searching
over more than a two dimensional parameter space, we will utilize     the well-known  Cyclic Jacobi Technique (CJT) for Hermitian matrix diagonalization, which is
based on two dimensional rotations. Then, we show
that for two dimensional rotations it is possible  to reduce the complexity of  the search  from a linear complexity
to a logarithmic complexity. The  proposed  algorithm is a blind realization of the CJT. Because of  the restriction to two dimensional rotations,    except for
the case of  \(n_{t}=2,\)   the OBNSLA  does not obtain
the null space after a finite amount of rotations, but rather  converges to the  null space  as the number of rotations increases.
Nevertheless, we will show (Sec \ref{Complexity and Convergence}) that
the OBNSLA  converges  to the null space very fast.
  We begin with a review the CJT.

\subsection{Review of the Cyclic Jacobi Technique}
\label{ReviewOfJacobiTechnique}

The CJT \citep[see e.g.][]{golubmatrix} obtains the  EVD of the \(n_{t}\times n_{t}\) Hermitian  matrix $\Gmat$ via a series of 2-dimensional rotations that eliminates two  off-diagonal elements at each step (indexed by $k$). It begins by setting     $\Amat_{0}
=\Gmat$ and then performs  the following  rotation operations \(\Amat_{k+1}= \Vmat_{k} \Amat_{k} \Vmat_{k}^*, \;k=1,2,...\),  where \beq\label{Rlm(theta,phi)}\Vmat_{k}=\Rmat _{l,m} (\theta,\phi)\eeq  is  an $n_{t} \times n_{t}$ unitary  rotation  matrix that is equal to \(\Imat_{n_{t}}\) except for its \(m\)th and \(l\)th diagonal entries that are equal to  \(\cos(\theta)\), and its \((m,l)\)th and \((l,m)\)th entries  that are equal to \(e^{-i\phi}\sin(\theta)\) and \(-e^{i\phi}\sin(\theta)\), respectively.
For each $k$, the values of $\theta,\phi$ are chosen such that
$[\Amat_{k}]_ {l,m}=0$,
 or stated differently, $\theta$ and $\phi $ are chosen to  zero the  $l,m$ and  $m,l$ off diagonal entries of $\Amat_{k}$ (which are conjugate to each other). Note that in an \(n_{t}\times n_{t}\) Hermitian matrix,  there are \((n_t-1)n_{t}/2\) such pairs.   The  values of $l,m$ are chosen in step $k$ according to a function
 $J:{\mathbb N}\longrightarrow \{1,...,n _{t}\} \times \{1,...,n_{t}\}, $    i.e $J_k=(l_{k}, m_{k})$. It is the choice of $J_{k}$
  that differs between different Jacobi techniques.   In the   cyclic Jacobi technique, $l_{k},m_{k}$  satisfy   $1<l_{k}<n_t-1$ and $l_{k}<m_{k}\leq n_{t}$ such that each pair $(l,m)$ is chosen once in every      \((n_t-1)n_{t}/2\)  rotations. Such \((n_{t}-1)n_{t}/2\) rotations are  referred  to as a  Jacobi sweep. An example of a single sweep   of the CJT for   $n_{t}=3$ is the following series of rotations:   $J_1 =(1,2) ,\;J_2=(1,3),J_3 =(2,3)$. The next sweep is $\;
 J_{4}=(1,2), J_{5}=(1,3),J_{6}=(2,3)$ and so forth.

The convergence of the CJT has been studied extensively over  the last sixty years.  
The first proof of convergence of the CJT for complex Hermitian matrices was given in \citep{forsythe1960cyclic}. However, this result did not determine the  convergence rate. The convergence rate problem was addressed  in \citep{ henrici1968estimate}, which proved that the CJT for real symmetric matrices has a global linear convergence rate\footnote{A sequence \(a_{n}\) is said to have a linear convergence  rate of \(0<\beta<1\) if there exists \(n_0\in \nat\) such that  \(\vert a_{n+1}\vert<\beta \vert a_{n}\vert \) for every \(n>n_0\). If \(n_{0}=1\), \(a_{n}\) has  a global linear convergence rate.}  if  $\theta_{k}\in[-\pi/4,\pi/4]$ for every $k$.    This result was extended to complex Hermitian  matrices in  \citep{FernandoNumerical1989}.
  It was later shown in \citep{HenficSpeed1958,WilkinsonNotes1962}  that  for  a matrix with well separated eigenvalues, the CJT has a  quadratic convergence rate\footnote{A sequence is said to have a quadratic convergence  rate if there exists  \(\beta>0,n_0\in\nat\) such that \(\vert a_{n+1}\vert<\beta \vert a_{n} \vert^{2}, \forall  n>n_0\).}. This result was  extended in \citep{VjeranSharp1991} to a  more general case which includes   identical eigenvalues and  clusters of  eigenvalues (that is, very close eigenvalues).   Studies have shown that in practice the number of iterations  that is required for the CJT  to reach its asymptotic quadratic convergence rate is a small number, but this has not been proven rigorously.  In \citep{BrentSolution1985} it is  argued heuristically that this number is \(O(\log_{2}(n_{t}))\) cycles for  \(n_t\times n_{t}\)  matrices. Extensive numerical results show that   quadratic  convergence  is obtained   after three to four cycles (see e.g.   \citep[page 429]{golubmatrix}, \citep[page 197]{parlett1998symmetric}).  Thus,  since each Jacobi sweep has \(n_{t}(n_{t}-1)/2\) rotations,   the overall number of rotations in the CJT roughly grows   as $n_{t}^{2}$.  For further details about the CJT and its convergence, the reader is referred  to \citep{golubmatrix,parlett1998symmetric}.

\subsection{  The One-Bit Line Search}
\label{ReveiwOfTheBNSLA}

   The  learning   in the  OBNSLA  is carried out in learning stages, indexed by \(k\),  where each stage performers one Jacobi rotation. The SU  approximates the matrix \(\Vmat\)  by  \(\Wmat_{k_s}\),  where \beq\label{DefineWk} \Wmat_{k}= \Wmat_{k-1}\Rmat _{l,m} (\theta_{ k},
 \phi_{k}),\;k=1,...,k_s,
\eeq  and $\Wmat_0=\Imat. $  Recall that in the CJT, one observes  the matrix \beq \label{definAkMinusOne} \Amat_{k-1}=\Wmat_{ k-1}^{*} \Gmat\Wmat_{k-1}\eeq and chooses \(\theta_{k}=\theta_{k}^{\rm J}\), \(\phi_{k}=\phi_{k}^{\rm J}\) such that
\beq\ [\Rmat^{*}_{l,m}(\theta_{k}^{\rm J },\phi^{\rm J}_{k})\Amat_{k-1} \Rmat_{l,m}(\theta_k^{\rm J},\phi_k^{\rm J} )]_{l,m}=0  \eeq
In the OBNSL  problem, the SU needs to perform this step using only \(\{\tilde\xvec (n),q_{} (n)\}_{n=1}\), without observing the matrix \(\Amat_{k-1}\).      The   following theorem, proved in \citep{Noam2012Blind},  is the first step towards such a blind  implementation of the Jacobi technique. The theorem  converts the problem of obtaining    the optimal Jacobi rotation angles into two one-dimensional  optimizations of the function \(S(\Amat_{k-1},\rvec_{l,m} (\theta,\phi))\) (which is continuous,   as shown in  \citep{Noam2012Blind}), where \(S(\Amat,\xvec)=\xvec^* \Amat \xvec\)  and   $\rvec_{l,m} (\theta,\phi)$ is $\Rmat_{l,m}(\theta, \phi)$'s $l$th column.

\begin{theorem}{\citep[Theorem 2]{Noam2012Blind}}
 \label{Theorem:TweDimensionalSearchToOneDemenssionalSearch}
Consider the $n_{t}\times n_{t}$  Hermitian matrix $\Amat_{ k-1}$ in \eqref{definAkMinusOne}, and let    $S(\Amat,\xvec)=\xvec^* \Amat \xvec$ and   $\rvec_{l,m} (\theta,\phi)$ be $\Rmat_{l,m}(\theta, \phi)$'s $l$th column. The optimal Jacobi parameters $\theta_{k}^{\rm J}$ and $\phi_{k}^{\rm J}$, which zero out the \((l, m)\)th entry of \( \Rmat^{*}_{l,m}( \theta_{k}^{\rm J},\phi^{\rm J}_{ k}) \Amat_{k-1} \Rmat_{l,m}(\theta_k^{ J},\phi_k^{\rm J})\),  are given  by
 \beqna
\label{ThetaOptimization}\phi_{k}^{\rm J} &=& \mathop {\arg\min  } \limits_{ \tiny \phi  \in[-\pi,\pi ]} S\left(\Amat_{k-1} , \rvec _{ l,m}(\pi/4, \phi)\right)\\ \label{EstThetaFromPhi} \theta^{\rm J}_k &=& T_{k}(\phi^{\rm J}_k) \eeqna where \beqna \label{optimal phi}
T_{k}(\phi)&=&\left\{ \begin{array} {lcl}  \tilde\theta_{k}( \phi) & {\rm if}\;- \frac \pi 4\leq \tilde\theta_{k}(\phi) \leq \frac \pi 4 \\\tilde\theta_{ k}( \phi)-{\rm sign}( \tilde \theta_{k }(\phi ))\pi/2 &{\rm otherwise}
\end{array} \right.\eeqna
where \({\rm sign}(x)=1\) if \(x>0\) and \(-1\) otherwise,
and \beq
\label{tilde phi Optimization}
\tilde\theta_{k}(\phi)= \arg \min_{ \theta\in[-\pi/2, \pi/2 ]} S\left( \Amat_{k-1}, \rvec_{ l,m} ( \theta, \phi) \right)
\eeq
\end{theorem}

  The theorem enables  the SU to solve the  optimization problems in \eqref{ThetaOptimization} and \eqref{tilde phi Optimization} via line searches  based on \(\{\tilde\xvec (n),q_{} (n)\}_{n=1}^{T}\). This is because under the OC, the SU can extract           \(\tilde h(\tilde \xvec(n),\tilde\xvec(n-m)),\)  which indicates whether  \( S(\Gmat, \tilde\xvec(n)) >S(\Gmat,\tilde\xvec(n-m))\) is true or false.
It is possible, however, to further  reduce the complexity of the line search, which is important, since each search point requires a TC.   To see this, consider the line search in \eqref{ThetaOptimization} and denote $w(\phi)=S(\Amat_k, \rvec_{l,m}(\pi/4,\phi))=\Vert\Hmat \Wmat_{k-1}\rvec_{l,m}(\pi/4,\phi)\Vert^{2}$. According to the OC, for each \(\phi_1,\phi_2\) the SU only knows whether \(w(\phi_1)\geq w(\phi_2)\) or not. Assume that the SU tries to approximate \(\phi^{\rm J}_{k}\)  by   searching over a linear  grid, with a spacing of \(\eta,\) on the interval \([-\pi,\pi]\).  The complexity of such a search is at least \(O(1/\eta)\) since each point in the grid must be compared to a different point at least once.      The two line searches  in  \eqref{ThetaOptimization} and  \eqref{tilde phi Optimization} would be    carried out much more  efficiently if binary searches could be invoked.
However,  a binary search is feasible only if the objective function has a unique local minimum point, which is not the case in  \eqref{ThetaOptimization} and  \eqref{tilde phi Optimization}  because \beq
\label{TrigonometricS} \begin{array}{lll}
S(\Gmat,\rvec_{l,m}(\theta, \phi))=\cos ^2(\theta ) \left| g_{l,l}\right| +\sin ^2(\theta ) \left|g_{m,m} \right| -\left|g_{ l,m}\right| \sin (2 \theta ) \cos ( \phi+\angle g_{l,m} )
\end{array}\eeq
 Thus,  before invoking  the binary search,   a single-minimum interval (SMI) must be determined; i.e, an interval in which the target function in \eqref{ThetaOptimization} or  \eqref{tilde phi Optimization} has a single local minimum. This is possible via the following proposition:
 \begin{proposition}   \label{OneBitAntisymetric} Let $w(\phi)=\Vert\Hmat \rvec_{l,m}(\pi/4,\phi)\Vert^{2}$ where $\rvec_{l,m}$ is defined in Theorem  \ref{Theorem:TweDimensionalSearchToOneDemenssionalSearch}. Let $\check\phi\in[- \pi,\pi]$ be a  minimum\footnote{Since $w$ is  a $2\pi$  periodical sinusoid, such a point always exists, though might not be unique.} point of $w(\phi),$ then
 \begin{enumerate}[(a) ]
  \item \label{c_a}    $\check \phi\in[-3\pi/4,-\pi/4]$ if $w(-\pi), w(0)\geq w_{}(-\pi/2)$.   \item      $\check \phi \in[-\pi/4,\pi/4]$ if $w(-\pi)\geq  w(-\pi/2)\geq w(0).$ \item     $\check \phi \in[\pi/4,3\pi/4]$ if $w(-\pi), w(0)\leq w(-\pi/2).$\item     $\check \phi \in[3\pi/4,\pi]\cup[-\pi,-3\pi/4]$ if $w(-\pi)\leq w(-\pi/2)\leq w_{}(0).$ \end{enumerate}  \end{proposition}
\IEEEproof From \eqref{TrigonometricS}, \(w(\phi)\) can be expressed as  $w(\phi)=B-A\cos( \phi-\check \phi) , \;A,B\geq0 $. If \(B=0\), every \(\phi\in[-\pi,\pi]\) is a minimum point and the proposition is true. We now assume that \(B>0\). By substituting  \(w(0)=B-A\cos(\check \phi), w(\pi/2)=B-A\sin(\check \phi)\) and \(w(\pi)=B+A\cos(\check \phi)\) into \(w(-\pi), w(0)\geq w_{}(-\pi/2)\), one  obtains that the latter is equivalent  to $\pm\cos(\check \phi)>\sin(\check \phi)$  which is equivalent to \((-3\pi/4<\check \phi<\pi/4)\cap((-\pi<\check \phi<-\pi/4)\cup(3\pi/4<\check \phi<\pi)).\) The last set can be written as  \(\check \phi\in(-3\pi/4,-\pi/4)\), which establishes \eqref{c_a}. The proof of (b)-(d) is similar.
 \hfill $\Box$

Note that unless  $w$ is a horizontal line, it is  a $2\pi$  periodical sinusoid. In the latter case,  there cannot be more than a single local (and therefor global) minimum within an interval of \(\pi/2\). If \(w\) is a horizontal line, every point is a minimum point. In both cases   the SU can efficiently approximate  \(\theta^{\rm J}_{k}, \phi^{\rm J}_{k}\) by \(\hat \theta_{k}^{\rm J}\) and \( \hat\phi _{k}^{ \rm J}\), respectively, using a binary search,  such that
\beq\label{OtimalThetaAndPhi}
\vert\hat\theta^{\rm J}_{ k}-T_{k}(\hat \phi_{k}^{ \rm J})\vert\leq
\eta,
\vert\hat\phi^{\rm J}_{k}- \phi_{ k}^{\rm J}\vert \leq \eta,
\eeq  where $\eta>0$ determines the approximation accuracy.
The SU uses Proposition \ref{OneBitAntisymetric} to determine an SMI via  \(u_{n}(\pi,\pi/2)\) and $u_{n}(\pi/2,0)$, where \(u_{n}(\phi_n,\phi_{n-1})=\tilde h_{n}(\rvec_{l,m}(\pi/4,\phi_{n}),\) \(\rvec_{l,m} (\pi/4,\phi_{n-1})),\) and \(\tilde h\) is defined in \eqref{Define H tilde}.     The one-bit line search is given in Algorithm \ref{Algorithm:ModifedLineSearch}.
 In  determining the SMI, the one-bit line search   requires 3 TCs: two TCs  for $u_n(\pi,\pi/2),$  and one more for $u_n(\pi,0)$.  Given an SMI of length \(a\) and an accuracy of  $\eta> 0$, it takes    $\lfloor-\log_{2}( \eta/a)\rfloor+1$ search points  to obtain the minimum to within  that accuracy. In the search for \(\phi^{\rm J}_{k}\), \(a=\pi/2\), thus   $\hat \phi^{\rm J}_{k}\in[ \phi^{\rm J}_{k}-\eta ,\phi^{\rm J}_{k} +\eta]$  is obtained using \([-\log(2\eta/\pi)]+1\) TCs, plus the 3 TCs  required for determining  the SMI and 2 more TCs to compare the initial boundaries of the SMI. Then, $ \theta_{k}^{\rm J}$ can be approximated to within $\eta$   in the same way.

We conclude with a discussion of  parameter \(M\) in the OC. The proposed line search can obtain  $w(\phi)$'s  minimum even for \(M=1\). However, the number of TCs is lower if  $M$ is larger. Assume that the SU has obtained an SMI  \([\phi^{\min},\phi^{\max}]\).  It takes the SU two TCs  to determine whether \(  w(\phi_{\max})>w(\phi_{\min}) \), where, in the first TC, it transmits \(r_{l,m}(\pi/4,\phi_{\max})\) and measures \(q(1)\), and in the second TC,  it transmits \(r_{l,m}(\pi/4,\phi_{\min})\), and measures \(q(2)\).  If  the latter inequality  is  false,   \(\phi_{\max}\) is set as  \(\phi_{\max} =(\phi_{\min}+\phi_{\max})/2\). In the next phase of the binary search the SU  cannot use   \(q(1)\) and  it needs an additional TC to do so. In general,  $M>\lfloor-\log(\eta/\pi)+1\rfloor$  guarantees that each search point requires one TC.


\subsection{The OBNSLA}
\begin{figure}
\centering
{ \epsfig{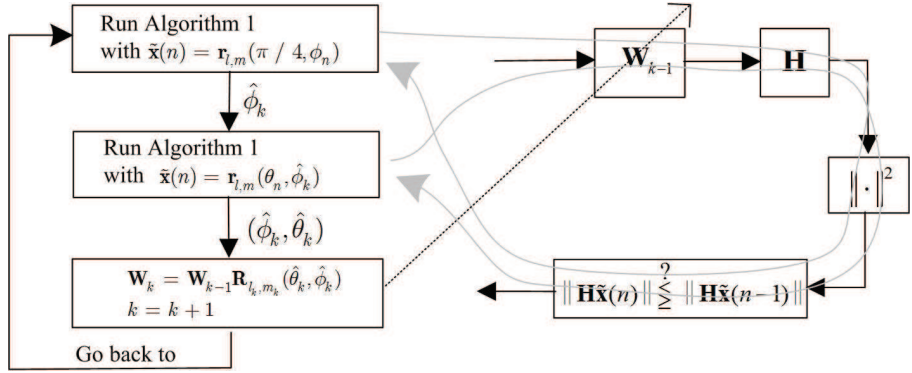}}
\caption{High-level description of the One-bit Blind Null Space Learning Algorithm. The gray arrows represent the path \(\tilde\xvec(n)\)
undergoes before \(q(n)\) is obtained by the SU. The dashed line
represents the action of  updating of the precoding matrix \(\Wmat_{k-1}\).}
\label{Figure:HighLevelDescription}
\end{figure}

Now that we have established the one-bit line search, we can present the OBNSLA.  In the OBNSLA, the  SU performs two line searches for each \(k\). The first  search is carried out to find \(\hat\phi_{k}\) that minimizes $\Vert\Hmat\Wmat_{k} \rvec_{l_{k}, m_{k}} (\pi/4, \phi))\Vert^{2}$, where  each search point, $\phi_{n}$, is obtained by one TC in which the SU  transmits       \beq \label{define x} \begin{array}{lll} \xvec_{s} (t)=\tilde \xvec(n)=\Wmat_{k} \rvec_{l_{k},m_{k}} (\pi/4, \phi_n)\in{ \mathbb C}^{n_{ t}} ~~~~~,\;\forall (n-1 )N \leq t\leq nN,\end{array} \eeq  and  measures  $q(n).$   In the first line search, the SU obtains \(\hat\phi^{\rm J }_{k}\) which is  then used in the second line search to obtain \(\tilde \theta_{k}(\hat\phi _{k}^{\rm J})\) according to \eqref{tilde phi Optimization}, and then to obtain \(\hat\theta^{\rm J}_{ k}\)   according to \eqref{EstThetaFromPhi}.
The indices \((l_{k},m_{k})\) are chosen  as in the CJT.
After performing \(k_s\) iterations the SU approximates  the matrix   \(\Vmat\) (see \eqref{EVD of G})  by \(\Wmat_{k_{s}}\). It then  chooses its   pre-coding matrix  \(\Tmat_{k_{s}}\) as
\beq
\label{DfinePks}
\Tmat_{k_{s}}=[\wvec^{k_{s}}_{ i_{1}},...,
\wvec^{k_{s}}_{i_{n_{t}-n_{r}}}],
\eeq
 where \(\wvec^{k}_{i}\) is \(\Wmat_{k}\)'s \(i\)th column, and  \(i_{1},i_{2 },...,i_{n_t}\) is an indexing such that
\(
(\wvec^{k_{s}}_{i_{q}})^{*}\Gmat \wvec^{s}_{i_{q}} \leq (\wvec^{ k_{s}}_{i_{v }})^{*}
\Gmat\wvec^{s}_{i_{v}}
\) for every \(q\leq v\).
Thus, the interference power that the SU inflicts  on the PU is bounded as \( \Vert \Hmat_{} \xvec_{} \Vert^{2}\leq p_{s}\Vert  \Hmat_{} \wvec^{k_{s}}_{i_{ n_{t}-n_{r}}}\Vert^{2},\forall \Vert\xvec\Vert^{2}=p_{s}, \) where $p_s$  is the SU's transmit power. A high level description
of the  OBNSLA algorithm is given in Fig. \ref{Figure:HighLevelDescription} and the exact
algorithm is given in Algorithm \ref{Algorithm:MBNSL}.

Although the SU becomes ``invisible'' to the PU after it learns  ${\cal N}(\Hmat_{ps})$, it interferes with the PU during this learning process. Furthermore, this interference is an important ingredient in the  learning since it provides the SU with the means to learn \({\cal N}(\Hmat_{ps})\), i.e. \(q(n)\). Nevertheless, the SU must also protect the PU  during the learning process. Hence we assume that there exists an additional mechanism  enabling the  SU to choose $\tilde\xvec(n)$'s power to be  high enough to be able to extract \(q(n),  \) but not too high, so as to meet the interference constraint  \eqref{InterferenceConstraint}. We give examples for such mechanisms   in \citep[Sec. II-C]{Noam2012Blind}.

\begin{algorithm}
\caption{$[z,n]={\rm \bf OneBitLineSearch} (\{\tilde h_{l}\}_{l\in\nat},z_{\max},n,\eta,\xvec(z))$}
\begin{algorithmic}\label{Algorithm:ModifedLineSearch}
\STATE
{\bf Initialize}:  $L\leftarrow z_{ \max }$,
\STATE $u_{n}(z_1,z_{2})\leftarrow \tilde h_{n}(\xvec(z_{1}),\xvec  (z_2))$

\STATE$a\leftarrow u_{n}(-L,-L/2), n++$.\STATE $b\leftarrow u_{n}( 0,-L/2 )$, \(n++\)
\STATE
{$z_{\max }\leftarrow (3+2 b-2 a (1+2 b) )L/4;$}
\STATE $z_{\min}\leftarrow z_{max}- L/2$.
\WHILE {$\vert z_{\max}- z_{ \min } \vert\geq \eta$}
\STATE {$z\leftarrow(z_{\max}+z_{ \min} ) /2$}\STATE
 $a\leftarrow u(z_{\max},z_{\min}), n++$
\IF{a=1}
\STATE $z_{\max}\leftarrow z$
 \ELSE \STATE $z_{\min} \leftarrow z$
\ENDIF
\ENDWHILE
\end{algorithmic}
\end{algorithm}

\begin{algorithm}
\caption{The OBNSL Algorithm}
\begin{algorithmic}  \label{Algorithm:MBNSL}

\STATE {\bf Input}: $\{\tilde h_{v}\}_{v\in {\mathbb N}}$, defined in \eqref{Define H tilde}.
\STATE {\bf Output}:\(\Wmat\)
\STATE {\bf initialize}: \(n=1\)
\STATE\([\Wmat,n]={\bf OBNSLF}(\{\tilde h_{v}\}_{v\in \nat},n_{t},n)\)
\STATE {End}
\end{algorithmic}
\begin{flushleft}{\bf Function}: \([\Wmat,n]\)= OBNSLF(\(\{\tilde h_{v}\}_{v\in \nat},n_{t},n\))\end{flushleft}
\begin{algorithmic}
\STATE {\bf Initialize}: $k=1$, $\Wmat=\Imat_{n_{ t}}$, $\Delta_{j}=2\eta, \forall j\leq 0$
\WHILE{$\left(\max_{j\in \{k-n_t(n_t-1)/2, ...,k\}} \Delta_{j}\geq \eta\right)$}
\STATE \(\xvec(\phi)\leftarrow\Wmat\rvec_{l_{k},m_{k}}(\pi/4,\phi)\)
\STATE \label{line:BNSL5}$[\hat \phi_{k},n] \leftarrow{\rm \bf OneBitLineSearch}\left(\{\tilde h_{l}\}_{l\in\nat},\pi,n,\eta,\xvec(\phi)\right)$\STATE \(\xvec(\theta)\leftarrow\Wmat\rvec_{l_{k},m_{k}}(\theta,\hat \phi_{k})\)
\STATE\label{line:BNSL6} $[\tilde\theta _{k},n] \leftarrow {\rm \bf OneBitLineSearch}\left(\{\tilde h_{l}\}_{l\in\nat},\pi/2,n,\eta,\xvec(\theta )\right)$
\STATE $\hat\theta_{k} \leftarrow\tilde \theta_{k}$ if $\tilde \theta_{k}\leq\vert \pi/4 \vert$, otherwise $\hat\theta_{k}\leftarrow\tilde \theta_{k}-\pi{\rm  sign}(\tilde \theta_{k})/2$.
\STATE  $\Delta_{k}\leftarrow \vert\theta_{k}\vert$
\STATE  $\Wmat\leftarrow \Wmat \Rmat_{l_{
k},m_k} (\hat\theta_k,\hat \phi_k)$
\STATE $k\leftarrow k+1$.
\ENDWHILE

\end{algorithmic}

\end{algorithm}

\section{Algorithm Convergence}
\label{Complexity and Convergence}
The OBNSLA is,  in fact, a blind implementation of the CJT whose convergence properties  have been extensively studied over the last 60 years. However,  the  convergence results  of the CJT do not  apply directly to the OBNSLA. This is because of the approximation in  \eqref{OtimalThetaAndPhi}; i.e., due to the fact that for every \(k\),  the rotation angles $\hat \theta_{k}^{\rm J}, \hat\phi_{k}^{\rm J}$ are obtained by a binary  search of  accuracy \(\eta \). Thus  the off diagonal entries are not completely annihilated; i.e.,   $[\Amat_{k+1}]_{l_{k},m_{ k}} \approx0$ instead of $[ \Amat_{k+1}]_{ l_{k},m_{k }}=0$.
  Moreover, we would like to make this line search accuracy as low as possible (that is, to make \(\eta \) as  large as possible) in order to reduce the number of TCs. It is therefore crucial to understand how  \(\eta \) affects the performance of the OBNSLA algorithm, in terms of convergence rate and   the interference reduction to the PU. In this section, we  extend the classic convergence  results  of the CJT to the    OBNSLA  and indicate the required accuracy in the binary search that assures convergence and bounds the   maximum  reduction level of the interference inflicted by the SU on the PU. It will also be shown that the same  convergence analysis applies to the BNSLA proposed in \cite{Noam2012Blind}.


 The following theorem shows that for a sufficiently good line search accuracy,   the OBNSLA  has a global linear   convergence rate.
\begin{theorem}\label{Proposition Linear Convergence} Let $\Gmat$ be a finite dimensional $n_{t }\times n_{t}$ complex Hermitian matrix and $P_{k}$ denotes the Frobenius norm of the off diagonal upper triangular (or lower triangular) part of $\Amat_{k}= \Wmat^{* }_{k-1 }\Gmat \Wmat_{ k-1}$ where \(\Wmat_{k }\) is defined in \eqref{DefineWk}    and let $m=n_{t}( n_{t}-1)/2.$
Let $\eta$ be the accuracy of the binary  search (see \eqref{OtimalThetaAndPhi}),  then the OBNSLA  satisfies \begin{equation}
 \label{EtaInequilityOfTheTheorem} \begin{array}{lll}P^{2}_{k+m} \leq P^{2}_{k}\left( 1-2^{- (n_{t }-2) (n_{t}-1) /2 } \right)+(n_{t }^{2 }-n_{t }) ( 7+2\sqrt 2 )\eta^{2 } \Vert \Gmat\Vert^{2}
\end{array} \end{equation}
  \end{theorem}
\IEEEproof See Appendix \ref{Appendix3}.

In   what follows, it will be  shown that for sufficiently small \(\eta\), the OBNSLA   has   an  asymptotic  quadratic convergence rate, but in order to obtain this,  we modify the   algorithm slightly  as follows. Let $I:\{1,...,n_{t}\}\rightarrow \{1,...,n_ {t}\}$ be the identity operator, i.e. $I(x)=x$.  At the beginning of each sweep, i.e.   for every   $k=q(n_{t }^{2}-n_{t})/2$ were $q\in \nat$, the SU sets $I_q=I $ and for each $k\in\{q(n_{t }^{ 2}-n_{t}) /2+1,...,(q+1 )(n_{ t}^{2}-n_{ t})/2\}$,  the SU modifies \(I_{q}\) as follows \beq \label{Eq26} I_q(l_{k})= \left\{
\begin{array}{l}l_{k}\text{ if } a_{l_{k}l_{k}}\geq a_{m_{ k},m_{k}}\\ m_{k} \text{   otherwise}\end{array} \right., \eeq where \(a_{l,m}\)
is the \((l,m)\)th entry of \(\Amat_{k}\) (defined in \eqref{definAkMinusOne}),
and \((l_{k},m_{k})\) are  determined by the rotation function
  \(J_{k}=(l_{k},m_{k})\)
of the CJT, as discussed in Section \ref{ReviewOfJacobiTechnique}. At the end of each sweep, i.e. for $k=(q+1)(n_{t }^{ 2} - n_{t} )/2$, the SU permutes the columns of $\Wmat_{k}$ such that  $\Wmat_{k}$'s $ l$th column becomes its $I_{ q}(l)$'s column. Note that this modification does not require extra TCs and that the  convergence result in Theorem \ref{Proposition Linear Convergence}  is  still valid. We refer to the OBNSLA  after this modification as the modified OBNSLA.

Besides the fact that this modification  is necessary for guaranteeing the quadratic convergence rate, as will be shown in the following theorem, it will also be shown that           it helps the SU to identify the null space (the last $n_{r }$ columns of $\Wmat_{k}$) blindly without taking extra measurements.

\begin{theorem}
\label{TheoremQuadraticConvergence}
Let $\eta$ be the accuracy of the binary  search,   \(\{ \lambda_{l} \}_{l=  1}^{ n_{ t}}\) be
  \(\Gmat\)'s eigenvalues  and let
  \beq
  \delta=\min_{\lambda_{ l} \neq \lambda_{r }} \vert \lambda_{ l} -\lambda_{r} \vert/3
  \eeq
    Let  $P_{k}$ be  the Fro{benius} norm of the off diagonal upper triangular  part of $\Amat_{ k}= \Wmat^{*}_{k-1} \Gmat \Wmat_{k-1}$, where \(\Wmat_{ k}\) is defined in \eqref{DefineWk}  and let $m=(n_{ t}^{2}-n_{t} )/2$.
Assume that the modified OBNSLA  has reached a stage \(k\), such that  \(P^{2}_{k}<\delta^{2}/8\), then
 \begin{equation}
 \label{EtaInequilityOfTheTheorem}
 \begin{array}{lll} P^{2}_{k+2m}\leq O \left(\left( \frac {P_{k+m }^{2 }}{ \delta} \right)^{2} \right)+O \left( \frac {  \eta P_{k+m}^{3/2 }}{\delta} \right) +O \left( \frac {  \eta ^{2}P_{k+m}^{1/2 }}{ \delta} \right) + 2\left(n_t^{2}-n_t \right) \eta^{2}\Vert \Gmat \Vert^{2}
\end{array} \end{equation}
  \end{theorem}
 Furthermore, the last \(n_{t }-n_{r}\) columns of \(\Wmat_{k+ m}\) inflict minimum interference to the PU; i.e.,   \( \Vert \Hmat_{ ps}\wvec^{k+m}_{i}\Vert\leq \Vert \Hmat_{ps}\wvec_{j}^{ k+m}\Vert ,\; \forall   1\leq j \leq n_{r}<i\leq n_{t}\).
\IEEEproof  See Appendix
 \ref{ProofOfTheoremQaudraticConvergence}.

Theorem \ref{TheoremQuadraticConvergence} shows  that to guarantee the quadratic convergence  rate, the  accuracy, \(\eta\), should be    much smaller than \(P_{k}^{2}\); that is,   let \(k_{0}\) be an integer  such that \(P^{2}_{k_{ 0}}<\delta^{2}/8,\) then  \beq\begin{array}{l} P_{k_0+2m}\leq O\left(\left( \frac{P_{k_0+m}}
{\sqrt\delta}\right)^{2} \right) \end{array}\eeq
if   \(\eta<<P^{2}_{k_{0}}\). This implies that  once \(P_{ k}\) becomes very small  such that    \(P^{}_{k}=O(\eta)\),  one cannot guarantee that  \(P_{k+2m}\)  will be smaller than \(P^{2}_{k+m}\) since at $k+1$ it will be  \(O(\eta)\).

  The  asymptomatic quadratic convergence rate  of Theorem \ref{TheoremQuadraticConvergence}   is determined by \(1/\delta\) where \(3\delta\)  is the minimal gap between \(\Gmat\)'s eigenvalues.  In addition, the quadratic convergence rate takes effect only after \(P_{k}^{2}<\delta/8\). Such a condition   implies that if \(\delta\) is very small, it will take the modified OBNSLA many cycles to reach its quadratic convergence rate. This is problematic  since MIMO wireless channels may have very close singular values (recall that \(\Hmat_{12}\)'s square singular values are equal to  \(\Gmat \)'s first \(n_{r}\) eigenvalues). If we were using the optimal  Cyclic Jacobi technique (i.e. no errors because of finite line search accuracy)  this would  not have practical implications since a quadratic decrease in \(P_{ k}\), which is independent of \(\delta\),  occurs prior to the phase where \(P_{k}^{2}< \delta/8\)  \citep{VjeranSharp1991}. In the following theorem, we extend this result to the modified  OBNSLA. \begin{theorem}
 \label{TheoremQuadraticConvergenceClusters}
Let $\eta$ be the accuracy of the line search, \(\{\lambda_{ l}\}_{l=1 }^{n_{t}}\) be \( \Gmat\)'s eigenvalues such that there exists a cluster of eigenvalues; i.e., there exists  a subset  \(\{ \lambda_{i_{l}}\}_{l=1}^{v}\subset \{\lambda_{l}\}_{l}^{n_{t}}\) such that    \(\lambda_{i_{l}}= \lambda+\xi_{l}, $ for  $ l\in L_{2} =\{i_{1},...,i_{v}\},\) where  \(\sum_{l=1 }^{v}\xi_{l}=0\) and the rest of the non-equal eigenvalues  satisfy  \(\delta_{c}>16\sqrt{ \sum \xi_{l}^{2}}\), where
\beq
  3\delta_{c}=\min(\Lambda_{1 }\cup \Lambda_{2})
\eeq
\beq\begin{array}{lll} \Lambda_1=
\left\{
  \vert\lambda_{l}- \lambda_{r} \vert: l\in L\setminus L_{2}, \; \lambda_{l} \neq \lambda_{r} \right\} \\ \Lambda_{ 2}= \left\{
  \vert \lambda_{l}- \lambda \vert : l\in L\setminus L_{2}\right\}\\
L=\{1,\ldots,n_{t}\} \end{array}
  \eeq
Then, once the modified  OBNSLA  reaches a $k$
such that   \beq \label{ClustreCondition} 2\delta_{c}\sqrt{ \sum_{l\in L_{2}} \xi_{l}^{2}}\leq P^{2}_{k}\leq \delta^{2}_{c}/8\eeq it satisfies\beq\begin{array}{l}\label{quadraticReduction} P^{2}_{k+2m}\leq  O \left(\left( \frac {P_{k+m}}{\delta_{c}} \right)^{4}\right)+O \left(\left( \frac {\eta  P_{k+m}^{3}}{\delta_{c}} \right) \right) +O \left(\left( \frac {\eta^{2}  P_{k+m}}{\delta_{c}} \right) \right)+2\left(n_t^{2}-n_t \right) \eta^{2}\Vert\Gmat\Vert^{2} \end{array}\eeq\end{theorem}
\IEEEproof See Appendix
 \ref{Appendix:ProofOfClusters}.

  Theorem \ref{TheoremQuadraticConvergenceClusters} states that in the  presence of a single  eigenvalue cluster; i.e. \(\sqrt{\sum_{l}\xi_{l}^{2 }}<< \delta_{c}\), and if \(\eta_{k}=o(P_{k})\), the modified  OBNSLA has   four convergence regions: The first region is  \(P^{2}_{k}\geq \delta^{2}_{c}/8\),  the second is \(2\delta_{c}\sqrt{ \sum_{l} \xi_{l}^{2}}\leq P^{2}_{k}\leq \delta^{2}_{c}/8\), the third is \(\delta/8\leq P^{2}_{k}\leq 2\delta_{c}\sqrt{ \sum_{l} \xi_{l}^{2}}\) and the fourth is \(P_{k}^{2}\leq\min_{l}\xi_{l}^{2}/8\). In the first and the third regions, the modified OBNSLA has at least a linear convergence rate while in the second and fourth regions, it has a quadratic convergence rate. This means that from a practical point of view, a close cluster of eigenvalues; i.e.  \(\sqrt{\sum_{l}\xi_{l}^{2 }}/\delta_{c}<<1 \), is not a problem. This is because      once the algorithm enters the second convergence region; i.e.,   it reaches the stage \(k=k_2\) such that   \(2\delta_{c}\sqrt{\sum_{l} \xi_{l}^{2}}<<P_{k_{2}}^{2}\leq \delta^{}_{c}/8\), \(P_{k}\) will decrease quadratically until \(k=k_3\) such that \(P_{k_3}^{2}\leq { 2\delta_{c}\sqrt{\sum_{l} \xi_{l}^{2}}} \). But the latter inequality implies  that \(P_{k_{3}}
<<P_{k_{2}}\), a fact that guarantees a significant reduction \(P_{k}\); i.e., from $P_{k_{2}}$ to $P_{k_{3}}$ with a quadratic rate. Nevertheless, \(P_{k}\) will  eventually   decrease  quadratically  as \(P^{2}_{k}\) becomes smaller than \(\delta/8\) as required by Theorem
  \ref{TheoremQuadraticConvergence}.    This phenomenon is also a characteristic of the Cyclic Jacobi technique
\citep{VjeranSharp1991}.

 We now consider the maximum level of interference that the SU inflicts on  the PU. Our aim here is to relate the asymptotic behavior of     the maximum interference to that of     $P_{k}$, and to        obtain bounds on the maximum interference   as a  function of \(\eta\).
 We begin with the following  proposition:
\begin{proposition} \label{CorollaryInterference}
Let \(\Tmat_{k}\) be the SU's pre-coding matrix defined in \eqref{DfinePks},   $\tvec_{i}^{k}$ be its $i$th column, $Q=\{1,...,n_{t}-n_{r}\}$, and  $P_{k}$ be the norm of the off diagonal upper triangular (or lower triangular) part of $\Amat_{k}$ (where \(\Amat_{k}\) is defined in \eqref{definAkMinusOne}). Then
\beq\label{34_12_4_12}
\max_{q\in Q}\Vert\Hmat_{12}\tvec_{q }^{k}\Vert^{2}\leq 2 P_{k}^{2}
\eeq
\end{proposition}
\IEEEproof This is an immediate result of \citep[Corollary 6.3.4]{horn_and_johonson} which states that for every  eigenvalue \(\hat \lambda\) of  $\Bmat+\Emat$, where $\Bmat$ is an \(n_{t}\times n_{t}\) Hermitian  matrix with eigenvalues $\lambda_{i}, i=1,...,n_{t}$, there exists $\lambda_{i}$ such that $\vert \hat \lambda-\lambda_{i}\vert^{2}\leq \Vert \Emat\Vert^{2}$, where $\Vert \cdot\Vert$ is the Forbinus norm. Thus, if one expresses $\Amat_{k}$ as $\Amat_{k}=\Bmat+\Emat$, where $\Bmat={\rm diag}(\Amat_{k}), \Emat={\rm offdiag}(\Amat_{k})$, \eqref{34_12_4_12} follows.  \hfill $\Box$

Since the maximum interference to the PU, for a single column of \(\Tmat_{k}\), is bounded by $2P_{k}^{2}$ (from Proposition \ref{CorollaryInterference}), it follows that the maximum interference satisfies \(\Vert\Hmat_{12}\Tmat_{k}\Vert^{2}\leq 2(n_{t}-n_{r})P^{2}_{k}\). Thus, it is possible to   apply the results of  Theorems \ref{TheoremQuadraticConvergence} and \ref{TheoremQuadraticConvergenceClusters} and to bound the maximum interference. These bounds are valuable since they relate the asymptotic level of interference to the accuracy of the line search $\eta$ (which is determined by the SU), thus enabling the SU to  control the interference reduction to the PU.

Before obtaining the first  bound on the interference, we need the following corollary of Theorem \ref{Proposition Linear Convergence}:
\begin{corollary}\label{CorollaryLimSup}
\beq\label{BoundOnInterference}
\lim\sup_{k}P_{k}^{2}\leq  \frac{(n_{t}^{2}-n_{t})(7+2\sqrt 2 )\eta^{2}\Vert \Gmat\Vert^{2 }}{2^{-(n_{t}-2)(n_{t}-1)/2}}.
\eeq
\end{corollary}
\IEEEproof See Appendix \ref{Appendix:CorollaryLimSup}.

From Corollaries \ref{CorollaryInterference} and  \ref{CorollaryLimSup} we obtain the following bound:\beq\label{BoundOnInterferenceSup}
\lim\sup_{k} \max_{q\in Q}\Vert\Hmat_{12}\tvec_{q }^{k}\Vert^{2}\leq  \frac{2(n_{t}^{2}-n_{t})( 7+2\sqrt 2 )\eta^{2}\Vert \Gmat\Vert^{2}}{2^{- (n_{t} -2)(n_{t}-1)/2}}.
\eeq

We now derive a tighter bound than \eqref{BoundOnInterferenceSup}  which is valid only if the conditions of   Theorem
\ref{TheoremQuadraticConvergence} are satisfied; i.e., that the OBNSLA is replaced by the modified OBNSLA and that  there exists  $k$ such that $P_{k}^{2}<\delta^{2}/8$. In this case,  by combining  Proposition \ref{CorollaryInterference} and  Theorem \ref{TheoremQuadraticConvergence}, one obtains
\beq\label{BoundOnInterference}
\begin{array}{l} \max_{q\in Q}\Vert\Hmat_{12}\tvec_{q }^{k}\Vert^{2}\leq O \left( \frac {P_{k}^{2}}{\delta} \right)^{2}+O \left( \frac {  \eta P_{k}^{3/2}}{\delta} \right)+O \left( \frac {  \eta^{2} P_{k}^{1/2}}{\delta} \right)+ 2\left(n_t^{2}-n_t \right) \eta^{2}\Vert\Gmat\Vert^{2}
\end{array}\eeq
Furthermore, if \(P_{k}\) becomes sufficiently small such that \(\eta> P_{k}\),   the dominant term in the RHS of
   \eqref{BoundOnInterference} will be \(O(\eta^{2});\) i.e., we effectively have:
\beq\label{BoundingMaxInterference}
 \max_{q\in Q}\Vert\Hmat_{12}\tvec_{q }^{k}\Vert^{2}\leq 2\left(n_t^{2}-n_t \right) \eta^{2}\Vert\Gmat\Vert^{2}+O(\eta^{2.5})
\eeq
Thus, the parameter \(\eta\)  gives the SU  autonomous control on the  maximum interference to the PU. 

We conclude  with the following corollary, which extends the convergence analysis presented in this section to the BNSLA. \begin{corollary} Theorems \ref{Proposition Linear Convergence}, \ref{TheoremQuadraticConvergence}, \ref{TheoremQuadraticConvergenceClusters}, Proposition \ref{CorollaryInterference} and  Corollary \ref{CorollaryLimSup} apply to the BNSLA  presented in \citep{Noam2012Blind}.
\end{corollary}
\IEEEproof
The proofs of Theorems \ref{Proposition Linear Convergence}, \ref{TheoremQuadraticConvergence}, \ref{TheoremQuadraticConvergenceClusters}, Proposition \ref{CorollaryInterference} and  Corollary \ref{CorollaryLimSup}  rely on the fact that the only difference between the CJT and the OBNSLA is in the rotation angles.  In the OBNSLA, the CJT's rotation angles , \(\theta_{k}^{\rm J},\phi_{k}^{\rm J}\),  are approximated   according to \eqref{OtimalThetaAndPhi}. Furthermore, note that  the BNSLA and the OBNSLA are identical except for the way in which each algorithm determines its   SMI (which are not identical SMIs) before invoking the binary search. However,  \eqref{OtimalThetaAndPhi} is satisfied as long as each SMI contains the desired minimum point. Because the  latter is satisfied by both algorithms, as indicated by   Proposition \ref{OneBitAntisymetric} for the OBNSLA and by Proposition 3 in \citep{Noam2012Blind} for the BNSLA, Theorems \ref{Proposition Linear Convergence}, \ref{TheoremQuadraticConvergence}, \ref{TheoremQuadraticConvergenceClusters}, Proposition \ref{CorollaryInterference} and  Corollary \ref{CorollaryLimSup} apply to the BNSLA.  \hfill $\Box$

We conclude this section with a discussion of the effective interference channel between the SU-Tx and the PU-Rx
\citep{zhang2010cognitive}.  In many MIMO communication systems the PU-Rx applies a spatial decoding matrix \(\Bmat\) to its received signal,  e.g, \(\Bmat\) might be
a projection matrix into the column space of the PU's direct channel \(\Hmat_{11}\). In this case, the equivalent received  signal is  \(\tilde\yvec_{p}(t)=\Bmat\yvec_{s}(t)\). Thus,  the SU's effective interference to the PU-Rx is \(\Bmat\Hmat_{}\tilde \xvec(n)\), rather than \(\Hmat\tilde \xvec(n)\).  We now discuss the effect of \(\Bmat\) on the OBNSLA algorithm and on the  bounds in \eqref{BoundOnInterferenceSup} and  \eqref{BoundingMaxInterference}, in different cases. The first is the  case where the SU   extracts \(q(n)\) from the PU's SINR and the PU calculates its SINR based on the effective interference; i.e., \(\Bmat\Hmat_{ps}\tilde \xvec(n)\). In this case, \(q(n)\) will be a one-bit function of \(\Vert  \Bmat\Hmat_{}\tilde \xvec(n)\Vert\), rather than  \(\Vert \Hmat\tilde \xvec(n)\Vert\). Thus, the OBNSLA will converge to the null space of the effective interference channel\footnote{ The channel \(\Bmat\Hmat\) is a special case of the effective interference channel which is defined in \citep{zhang2010cognitive} for the case for a MIMO TDD PU, where the SU interferes with both the PU-Rx and the PU-Tx; i.e., uplink and downlink.}; i.e.,  \( {\cal N}(\Bmat\Hmat\))  and the  bounds in \eqref{BoundOnInterferenceSup} and  \eqref{BoundingMaxInterference} will be valid by replacing \( \Hmat\) with \(\Bmat\Hmat\) and setting \(\Gmat=\Hmat^{*}\Bmat^{*}\Bmat\Hmat\). Note that for the OBNSLA to work, the matrix \(\Bmat\) must be constant
during the learning process. If the PU modifies \(\Bmat\)  due to the modifications
in the SU's learning signal, the OBNSLA will not converge to the null space. The second case is where the SU observes \(q(n)\) which carries information only on  \(\Vert\Hmat\tilde \xvec(n)\Vert\), while the effective interference is \(\Vert\Bmat\Hmat\tilde\xvec(n)\Vert\). In this case, the  OBNSLA will converge to \({\cal N}(\Hmat)\), but the bounds will not necessarily hold. If \(\Bmat\) is a projection matrix; i.e., it projects to some subspace of \(\comp^{n_{r}}\)  and it does not amplify the signal, the bounds  in \eqref{BoundOnInterferenceSup} and  \eqref{BoundingMaxInterference} will be satisfied. The SU can lose degrees of freedom  by restricting its signal to \({\cal N}(\Hmat)\) rather than \({\cal N}(\Bmat\Hmat)\) because \({\cal N}(\Hmat)\subseteq{\cal N}(\Bmat\Hmat)\).

\section{simulations}\label{Section:Simulation}
In this section we study the performance of the OBNSLA via simulations. We first compare the OBNSLA to the OBNSLA \citep{Noam2012Blind} in the non-asymptotic regime. We examine the effect of important practical aspects, such as  time-varying environments and quantization. In the second part of this section, we compare the  asymptotic properties of the two algorithms with respect to the asymptotic analysis in Sec. \ref{Complexity and Convergence}.
\subsection{Non-Asymptotic Comparison   of the BNSLA and the  OBNSLA}
We now  examine the OBNSLA and BNSLA in time-varying environments where  the SU  measures the PU's  transmitted signal and  extracts \(q(n)\) from it. Recall that it takes the SU some time to learn the null space of the interference channel. Thus, if the channel varies faster than the SU's learning period, it will not be able to effectively mitigate interference to the PU. Another problem that we address in this section is the effect of the variations in the PU  direct link; i.e., \(\Hmat_{pp}\),  on the performance of the OBNSLA and the BNSLA. We   assume that the primary user performs a power adaptation every 1 msec and that the SU sets this as the TC's length; the considerations for this choice are the following. Let \(T_p\) be the time  cycle  in which the PU performs  power control, \(T_{\Hmat_{pp}}\) and \( T_{\Hmat_{ps}}\) be the coherence times of  \(\Hmat_{pp}(t)\)  and \(\Hmat_{ps}(t),\) respectively, and  \(T_{\rm TC}\) be the length of the TC (which is equal to $N$ time units, as depicted in Fig. \ref{Figure:index}). Note that \(T_{\rm TC}\geq T_{p}\). Because the OBNSLA is based on the OC,  the SU must choose \(T_{\rm TC}\) such  that \(T_{\rm TC}<< T_{\Hmat_{ps}}\), where the  variations in the SU's  signal must be  the dominant factor affecting   the PU's SINR. If the PU adapts its transmission scheme fast, such that  \(T_p\) is  smaller than  typical values of \(T_{\Hmat_{pp}}\) and \(T_{\Hmat_{sp}}\), the SU can set \(T_{\rm TC}=T_{p}\).   In    LTE, for example, the PU can adapt its signal's     power, modulation and coding every 1 msec\footnote{The power control in  LTE  is typically no more than a few hundred  Hz. \citep[see e.g.][Section. 20.3]{LTE2009FromTheoreyToPractice}. However,  in the future it may be possible to   have  power control every 1 msec, since this is the  length of a Transmission Time Interval (TTI), the smallest   interval  in which a base station can schedule any user for  transmission (uplink or downlink).}. Other examples are  all   the third generation cellular systems  which perform  power control every 2/3 msec (W-CDMA) or 5/4 msec (CDMA2000)\footnote{Note that CDMA channels are not necessarily narrow-band.  However, the SU can still learn the null space of a narrow-band portion   since the PU will perform power adaptation  in the presence of   narrow-band interference as long as it is not too narrow. For instance the bandwidths of  most
CDMA2000 systems is 1.25 MHz, but the coherence bandwidth  typically varies between 50  KHz and 3 MHz.
   }  (see e.g. \citep{goldsmith2005wc}, Appendix D).

\begin{figure}
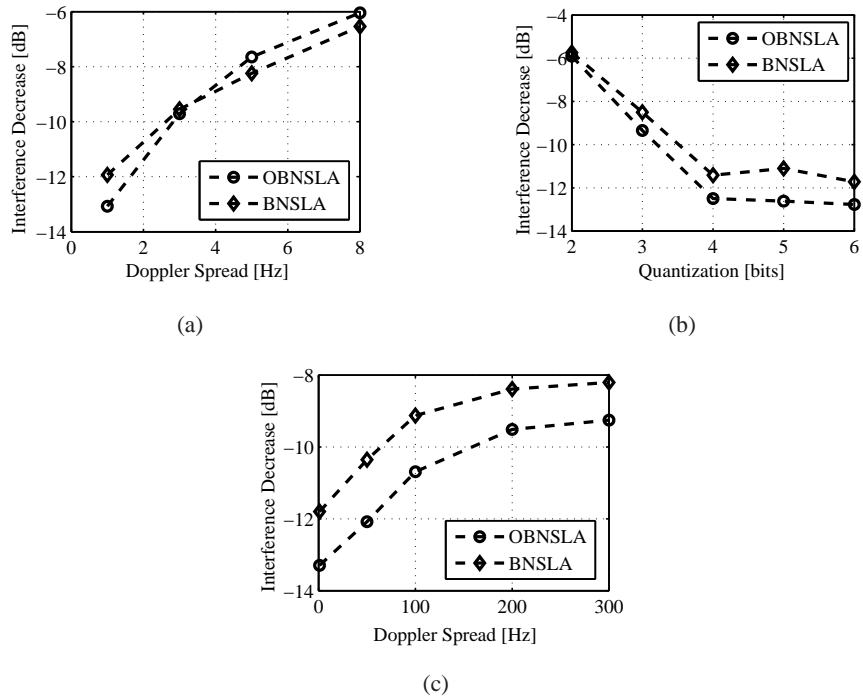

\centering
\subfigure[\label{Fig3(a)} ]
{  \epsfig{figure=BNSLVsOBNSLasDopller.eps, width=5.5cm}}
\qquad
\subfigure[\label{Fig3(b)}   ]{ \epsfig{figure=OBNSLA_vsBNSLA_quantization.eps,width=5.5 cm}}
\qquad
\subfigure[\label{Fig3(c)}   ]{ \epsfig{figure=OBNSLvsBNSLADifferentFpp.eps,width=5.5 cm}}
\caption{ Interference reduction after a single Jacobi sweep as a function of (a) \(\Hmat_{ps}\)'s Doppler spread, (b) SINR
quatization bits, (c) \(\Hmat_{pp}\)'s Doppler spread.    The maximum power constraint of each Tx is  23 dBm.      The channels'  path-losses  are calculated as  \(128.1+37.6\log_{10}(R)\), where \(R\) is the distance between the Rx and the TX in meters as used by the 3GPP (see page 61 3GPP Technical Report  36.814). The locations of the PU-TX  is randomly chosen from a uniform distribution over a 300 m  disk, and the locations of the  SU-Tx and the  SU-Rx are randomly chosen from a uniform distribution over  a 400 m disk. Both disks are  centered at the location of  the PU-Rx. The minimum distance between the PU-Rx to the  PU-Tx, and  the PU-Rx to the  SU-Tx is 20 m and 100 m, respectively.  For each \(t\) the entries of the channel matrices  \( \Hmat_{pp}(t),\Hmat_{sp}(t)\) and \(\Hmat_{ps}(t)\) are  i.i.d. where each entry is 15 KHz    flat fading Rayleigh  channel.  In all plots, the  Doppler spread of \(\Hmat_{sp}(t)\) is  15 Hz, which is also  \(\Hmat_{pp}\)'s
 Doppler
spread in (a) and (b). The Doppler spread of \(\Hmat_{ps}\) in
(b) and (c) is 1 Hz.  All channels are generated using the Improved Rayleigh Fading Channel Simulator \citep{XiaoNovel2006}.   The noise level at the receivers is -121 dBm and the SU transmit power during the learning process is 5 dBm.  The numbers of antennas are   \(n_{t_{s}}=3,\)
\(n_{t_{p}}=2\), \(n_{r_{p}}=1\) \(n_{r_{s}}=2\). }
\label{Figure:OBNSLA vs BNSLA}
\end{figure}

 Figure \ref{Figure:OBNSLA vs BNSLA} presents simulation results
where   the PU performs a power adaptation to  maintain a target 10 dB SINR at the receiver, and the SU inflicts interference on the PU and measures \(q(n)\) by listening to the PU signal's power at the SU-Rx\footnote{The SU sets \(q(n)=\frac 1 N\sum_{t=Nn}^{nN+N'-1}\Vert\yvec_{s}(t)-\bar\yvec_{s}\Vert^{2}\), where \(\bar \yvec_s\) is the average of \(\bar\yvec_{s}(t)\) over $t=N(n-1)+N',\ldots Nn+N'-1.$ The consideration for choosing such \(q\) are described in   \citep[Sec II-B1]{Noam2012Blind}.}. Fig. \ref{Fig3(a)}. presents the interference reduction of the BNSLA and the OBNSLA as a function of \(\Hmat_{ps}\)'s Doppler spread.
The results show     that both algorithms perform similarly but the BNSLA shows a slightly   better interference reduction than the  OBNSLA for higher values of Doppler spread and vice-versa.     { The fact that the BNSLA is slightly better than the OBNSLA in low Doppler spread can be explained by the fact that the BNSLA is a little  faster than the OBNSLA, as  we  discuss below. In  \citep[Proposition 3]{Noam2012Blind} it was shown that the BNSLA determines a   \(\pi/4\)-length SMI  using  \(w_{k}(0),w_{k}(\pi/2),  w_{k}(-\pi/2)\) if   \beq\label{ConditionCeck} \vert w_{k}(0)-w_{k}(\pi/2)
\vert\geq\vert w_{k}(\pi/2)-w_{k}(\pi)\vert, \eeq  where \(w_{k}(\phi)\) is some monotone function of the PU SINR such that obtaining its value for a given \(\phi\) requires a TC.  Thus, if  \eqref{ConditionCeck} is true, the BNSLA is faster  than the OBNSLA by one TC. This is due to the fact that   although  both algorithms require three TCs to determine an SMI, the BNSLA's SMI is half the  length of the OBNSLA SMI. If  \eqref{ConditionCeck} is false, the SU will determine  a $\pi/4$-length SMI by testing a similar condition to \eqref{ConditionCeck}  using \(w_k(0),w_k(-\pi/2), w_k(-\pi)\). This  requires one more TC to obtain \(w(-\pi)\). The same phenomenon occurs in the search for the \(\theta_{k}^{\rm J}.\) On the other hand, the inequality in  \eqref{ConditionCeck}  also suggests   a possible  explanation for the slightly better performance of the OBNSLA for low Doppler spread. The fact that the BNSLA, in order to determine an SMI, must check a condition that involves three noisy search points, while the OBNSLA does so via two noisy search points, which makes the OBNSLA more robust to noise. In low Doppler spread, this advantage  compensates for the fact that the OBNSLA  is slightly slower than the BNSLA. However, as the Doppler spread increases, the BNLSA's interference reduction becomes equal to that  of the OBNSLA and eventually becomes better.}

An important practical issue in the implementation the OBNSLA and the BNSLA is the granularity in the PU's SINR, which prevents  the SU from detecting small variations in the PU SINR.  A full theoretical convergence analysis of this problem is an important topic for future research. In this paper, we test  this problem using  simulation. Fig. \ref{Fig3(b)} presents simulation results for a scenario where  the PU's power control process is based on a quantized measurement of the  PU's SINR in the range $-5$ dB to $20$ dB. It is shown that  the interference reduction is not improved for more than 4 bits of quantization. This means that small granularity does have a practical  affect on the performance of the OBNSLA and BNSLA.

In the last simulation we investigate the effect
of variations in the PU direct channel on the performance of
the OBNSLA and BNSLA. Recall that the OC requires that the  PU's SINR be mostly
affected by the variations in the interference inflicted by the SU. However, if
the PU direct channel varies, it will be impossible for the SU to distinguish
whether the variations in the PU SINR are due to the SU interference
signal or to the PU's direct channel path loss, and will lead to errors.   To study this
phenomenon we run a simulation for the  case
where the PU experiences fast fading. Fig. \ref{Fig3(c)}
presents the interference reduction of the two algorithms as a function
of \(\Hmat_{pp}\)'s Doppler spread, where \(\Hmat_{pp}\) is Rayleigh
fading. The result shows that OBNSLA outperforms the
BNSLA at all frequencies. Even for
a 150 Hz doppler spread, the OBNSLA achieves a 10 dB interference
reduction.

\subsection{Asymptotic analysis}
\begin{figure}
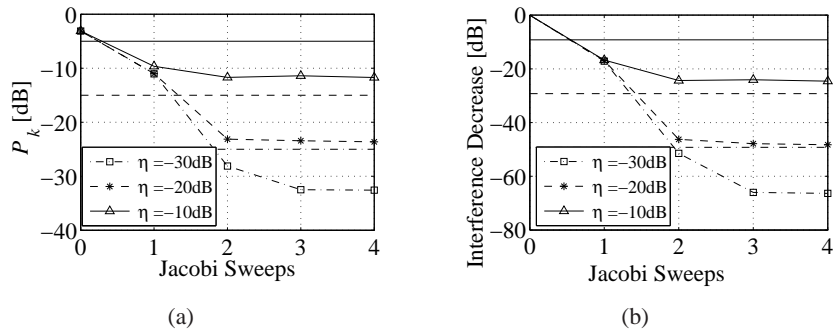

\centering
\subfigure[\label{Fig4(a)} ]
{ \epsfig{figure=SumOfOffDiag.eps, width=5cm}}
\qquad
\subfigure[\label{Fig4(b)}  ]{ \epsfig{figure=InterferencePlotWithBound.eps, width=5cm}}
\caption{Simulation results for different values \(\eta\)  of the OBNSL algorithm to  obtain the null space of \(\Hmat\) with \(n_{t}=3\) transmitting antennas and \(n_{r}=2\) antennas at the PU receiver. The matrix \(\Gmat=\Hmat^{*}\Hmat \) was  normalized such that \(\Vert\Gmat \Vert^{2}=1\).
The unmarked lines in (a) and (b) represent the asymptotic upper bound of \eqref{BoundOnInterferenceSup} and  \eqref{BoundingMaxInterference} respectively on the corresponding marked line; e.g. the solid unmarked line is a bound on the solid line with squares.   We used 200 Monte-Carlo trials where the entries of  \(\Hmat\)   are i.i.d. complex Gaussian random variables.}
\label{Figure:Convergence}
\end{figure}

We now compare the asymptotic properties of the OBNSLA to the bounds derived in Section \ref{Complexity and Convergence}  under optimal conditions; i.e, \(q(n)\) is perfectly observed. Figure \ref{Fig4(a)} depicts $P_k$ and the bound on it as given in \eqref{BoundOnInterferenceSup}, versus complete OBNSLA sweeps; i.e. \((n_{t}^{2} -n_{t})/2\) learning phases.   It shows that for sufficiently small \(\eta\) the   OBNSLA converges quadratically. The quadratic decrease   breaks down when the  value  of \(P_{k}\) becomes as small as the order of magnitude of    \(\eta\). This result is consistent  with    Theorem \ref{TheoremQuadraticConvergence}. 
   Figure \ref{Fig4(b)} depicts the interference decrease and the bound on it as  given in \eqref{BoundingMaxInterference} versus  the number of transmission  cycles. It  shows that    the asymptotic level of the interference to the PU is  \(O(\eta^{2})\).
Note that because the simulations in  Fig. \ref{Figure:Convergence} are under optimal conditions; i.e., \(q(n)\) is not noisy and the channel is not time-varying, we would obtain the same results  if we applied the BNSLA instead of the OBNSLA. In what follows, we study asymptotic performance under non-optimal conditions.

The bounds in this paper are derived under the assumption that   the OC holds perfectly. In practice, however,  \(q(n)\) is affected by measurement error such as noise.  Furthermore, the channel matrices${{\mathbf \ }{\mathbf H}}_{pp}{\mathbf ,\ }{{\mathbf H}}_{sp}{\mathbf \ }$\textbf{ }vary with time, a fact that may also affect the function $q(n)$. For example, variations in $\Hmat_{pp}$  affect the PU's SINR.  In addition,  $\Hmat_{ps}$ is also time-varying, which leads to some discrepancy between the estimated null space and the true null space. In what follows, we show in simulations  that the derived bounds are still  useful in practice.   A full theoretical convergence analysis of the O/BNSLA in practical conditions, which extends   the bounds derived in this paper  to account for measurement noise and time variations in the channel is   a  topic for future research, beyond the scope of this paper.

\begin{figure}
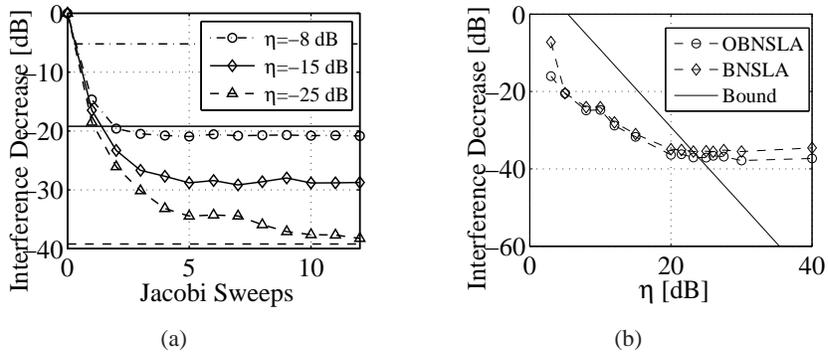

\centering
\subfigure[\label{Fig5(a)} ]
{ \epsfig{figure=convergeFig.eps, width=5cm}}
\qquad
\subfigure[\label{Fig5(b)}   ]{ \epsfig{figure=BoundVsPerformanc.eps, width=5cm}}
\caption{ Interference reduction (marked lines) of the OBNSLA, as a function of Jacobi Sweeps (a), and of the OBNSLA and BNSLA as a function of \(\eta\) (b).  \(\Hmat_{ps}\) was  generated according to \citep{liou2009characterization}, which represents a fixed Rx-Tx channel where one antenna is 1.75 m in height  and the second antenna is 25 m in height. The Rician factor and the Doppler spread of  \(\Hmat_{ps}\)  were determined according to Equations (13) and (14) in \citep{liou2009characterization}.   The unmarked lines represent the  bound in \eqref{BoundingMaxInterference} for the corresponding   marked lines with the same pattern, e.g., in Subfigure (a) the dotted-dashed unmarked line represent the bound on the OBNSLA's interference reduction   with \(\eta=-8\) dB, which is represented by the dotted-dashed line that is marked with circles. The numbers of antennas are   \(n_{t_{s}}=2\)
\(n_{t_{p}}=2\), \(n_{r_{p}}=1\) \(n_{r_{s}}=2\). The results were averaged over  1000 Monte-Carlo trials.}
\label{Figure:FixedSUPUConvergence}
\end{figure}

Figure \ref{Figure:FixedSUPUConvergence} presents simulation  results for an identical scenario as in Figure \ref{Figure:OBNSLA vs BNSLA} except for      \(\Hmat_{ps}\) which  is generated assuming that both the SU-Tx and PU-Rx are fixed.     The result shows that  for an interference  reduction  smaller or equal to 37 dB,    the bound in \eqref{BoundingMaxInterference} predicts the behavior of the  interference reduction (i.e.,  it     decreases as \(\eta^{2}\)) of both the  OBNSLA and the BNSLA. Furthermore, it is shown that the BNSLA and the OBNSLA  have the same asymptotic properties; i.e. convergence rate and asymptotic interference reduction.

\section{Summary and Future Research}
\label{Section:coclusions}

 This paper proposed  the OBNSLA, which enables a MIMO CR SU to learn the null space of the interference channel to the PU  by observing a binary function that indicates  the variations (increase or decrease) in the PU's SINR. Such information can be extracted,  for instance,  from the quantized version of the variation in the PU's SINR, or in the  PU's modulation. We also provided a  convergence analysis of the OBNSLA, which also applies to the BNSLA \citep{Noam2012Blind}. It was shown that  the two algorithms have a  global linear and an asymptotically quadratic convergence rate. It was also  shown  in simulations  that just like in the  Cyclic Jacobi technique, the OBNSLA and the BNSLA reach their  quadratic convergence rates in only  three to four cycles. In addition, we derived asymptotic  bounds on the maximum level of interference that the SU inflicts on the PU.  The derived  bounds  have  important practical implications. Due to the fact that  these   bounds  are   functions of a parameter  determined by the SU,  it enables the  SU  to control  the maximum level of interference caused to the PU. This gives the OBNSLA (or the BNSLA) a useful stopping criterion which guarantees the   protection of  the PU.
 The analytical convergence rates and interference bounds were validated by extensive simulations.

We consider the theoretical analysis of the OBNSLA and BNSLA under measurement noise as an important  topic for future research. Note that in the presence of noise, the analysis of the two algorithms is not identical since the BNSLA relies on a continuous-valued function of the PU's SINR, whereas the OBNSLA relies on a binary function.
Noise, which is continuous-valued, will thus affect these functions and hence the performance and convergence of the two algorithms quite differently.\appendices
\section{}
\label{Appendix3}
Consider the first sweep of  the  BNSL algorithm; i.e.  $k=1,2,..., n_{t}(n_{t}-1)/2$.   Denote the number of rotated elements in the $l$th row by  \(b_{l}=n_{t}-l\) and let \beq \label{DefineW(l,k)}
\begin{array}{lll} c_{l} =\sum_{j=1}^{l} b_{j}=(2n_{t}-1-l)l/2;~ Z(l,k)= \sum_{j=1 }^{n_{t}-l}\vert [\Amat_{ k}]_{j+l,l}\vert^{2} ;~
W(l,k)=\sum_{j=l+1}^{n_{t}-1}  Z(j,k)
\end{array}\eeq
Note that \(W(0,k)=P^{2}_{k}\).
In every sweep, each entry is eliminated once; we therefore  denote \(\Amat_{k}\)'s $p,q$ entry  before its annihilation  as  $g_{q,p}(t)$ where $t$ denotes the number of changes since  $k=0$. After $g_{q,p}(t)$ is annihilated once, it will be denoted by $\tilde g_{q,p}(\tilde t)$ where $\tilde t$ is the number of changes after the annihilation. The   diagonal entries of $\Amat_{k}$ will be denoted by $x$   since we are  not interested in their values in the course of the proof. This is illustrated in the following example of a $4\times 4$ matrix
\beq\begin{array}{cc}
\begin{array}{c}
\Amat_0=
\left(
\begin{footnotesize}\begin{array}{cccc}
 g_{1,1}(0) & g_{1,2}(0) & g_{1,3}(0) & g_{1,4}(0) \\
 g_{2,1}(0) & g_{2,2}(0) & g_{2,3}(0) & g_{2,4}(0) \\
 g_{3,1}(0) & g_{3,2}(0) & g_{3,3}(0) & g_{3,4}(0) \\
 g_{4,1}(0) & g_{4,2}(0) & g_{4,3}(0) & g_{4,4}(0)
\end{array}\end{footnotesize}
\right), \\
\end{array} & \begin{array}{c}
\Amat_1 =
\left(
\begin{footnotesize}\begin{array}{cccc}
 x_{} & \epsilon  & g_{1,3}(1) & g_{1,4}(1) \\
 \epsilon  & x_{} & g_{2,3}(1) & g_{2,4}(1) \\
 g_{3,1}(1) & g_{3,2}(1) & x_{} & g_{3,4}(0) \\
 g_{4,1}(1) & g_{4,2}(1) & g_{4,3}(0) & x_{}
\end{array}\end{footnotesize}
\right), \\
\end{array} \\
\begin{array}{c}
~\Amat_2=
\left(
\begin{footnotesize}\begin{array}{cccc}
 x & \tilde{g}_{1,2}(0) & \epsilon  & g_{1,4}(2) \\
 \tilde{g}_{2,1}(0) & x & g_{2,3 }(2) & g_{2,4}(1) \\
 \epsilon  & g_{3,2}(2) & x & g_{ 3,4}(1) \\
 g_{4,1}(2) & g_{4,2}(1) & g_{4, 3}(1) & x
\end{array}\end{footnotesize}
\right), \\
\end{array} & \begin{array}{c}
~\Amat_3 =
\left(
\begin{footnotesize}\begin{array}{cccc}
 x & \tilde{g}_{1,2}(1) & \tilde{g}_{1 ,3}(0) & \epsilon  \\
 \tilde{g}_{2,1}(1) & x & g_{2,3 }(2) & g_{2,4}(2) \\
 \tilde{g}_{3,1}(0) & g_{3,2}(2) & x & g_{3,4}(2) \\
 \epsilon  & g_{4,2}(2) & g_{4,3}(2) & x
\end{array}\end{footnotesize}
\right) \\
\end{array} \\
\end{array}
\eeq
For arbitrary \(n_t\), after the first $c_{1}$ sweeps $\Amat_{c_{1}}$'s  first column is equal to the following vector:
\beq
[x,\tilde g_{2,1}(n_{t}-3),...,\tilde g_{n_{t}-1,1}(0),\epsilon_{c_1}]^{\rm T}
\eeq
and \beq Z(1,c_{1})\leq\vert \tilde g_{2,1}(n_{t}-3) \vert^{2}+...+\vert \tilde g_{n_{t}-1,1}(0)\vert^{2}+\vert \epsilon_{c_1} \vert^{2}
\eeq
From \eqref{definAkMinusOne} it follows that  for $q=2,...,n_{t}$,
\beq
\begin{array}{rll}
\tilde g_{q,1}(n_{t}-q-1)&=&\cos \left(\theta _{n_{t}-1}\right) \tilde{g}_{q,1}(n-q-2)-e^{i \phi _{n_{t}-1}} g_{q,n_{t}}(1) \sin \left(\theta _{n_{t}-1}\right)\\ ~&\vdots&~ \\\tilde g_{q,1}(1)&= &\cos \left(\theta _{q+1}\right) \tilde{g}_{q,1}(0)-e^{i \phi _{3}} g_{q,q+2}(1) \sin \left(\theta _{3}\right)
\\
\tilde g_{q,1}(0)&=&\epsilon_{q-1}  \cos \left(\theta _q\right)-e^{i \phi _q} g_{q,q+1} (1)\sin \left(\theta q\right)
\end{array}
\eeq
 where $\tilde g_{q,1}(-1)=\epsilon_1$.   The following bounds on $\{ \tilde g_{q,1}(l)\}_{l=0}^{n_{t}-q-1}$ are  obtained recursively (i.e., by  obtaining a bound on  $\tilde g_{q,1}(0),$ substituting and obtaining a bound on $\tilde g_{q,1}(1)$ and so on)
\beq\begin{array}{lll}\label{tilde g}
\tilde{g}_{q,1}(n_{t}-q-1)\leq\vert  \epsilon_{q-1}  \prod _{v=q}^{n_{t}-1} \cos \left(\theta _v\right)-\sum _{j=q}^{n_{t}-1} e^{i \phi _j} \sin \left(\theta _j\right) g_{q,j+1}(1) \prod _{v=j+1}^{n_{t}-1} \cos \left(\theta _v\right)\vert\\~~~~~~~ ~~~~~~\leq\vert\vvec(q)^{\rm T}\yvec(q)\vert+ \epsilon \prod _{v=q}^{n_{t}-1} \cos \left(\theta _v\right)\end{array}
\eeq where  $\vvec,\yvec\in\mathbb{C}^{ n_{t}-q}$ such that  $[\vvec(q)]_{j}=e^{i \phi _{j+q-1}} g_{q,j+q}(1)$,  $[\yvec( q)]_{j}=\sin \left(\theta _{j+q-1}\right) \prod _{v=j+q}^{n_{t}-1} \cos \left(\theta _v\right)$, \(j=1,...,n_{t}-q\), and $\epsilon=\max_q \vert\epsilon_q\vert$.  It follows that
\beq\label{before proposition8}
\begin{array}{lll}
\vert \tilde{g}_{q,1}(n_{t}-q-1) \vert^{2} &\leq\vert\yvec^{\rm T}(q)\vvec(q)\vert^{ 2}+\vert \epsilon\vert^{2} \prod _{v=q}^{n_{t}-1} \cos^{2} \left(\theta _v\right) \leq \Vert\yvec(q)\Vert^{ 2}\Vert \vvec(q) \Vert^{2}+\vert \epsilon\vert^{2} \prod _{v=q}^{n_{t}-1} \cos^{2} \left( \theta _v\right)
\end{array}\eeq

\begin{proposition}
\label{proposition8}
\beq\label{InductionAssumption}\begin{array}{l}
\Vert\yvec(q)\Vert^{2}=1-\prod_{i=q }^{n-1} \cos(\theta_{i})
\end{array}\eeq
\IEEEproof
This is shown by induction. By definition \beq\label{yvecDefinition}\begin{array}{l}\Vert \yvec(q)\Vert^{2}= \sum_{i=q}^{ n-1} \sin^{2}(\theta_{i}) \prod_{v=i+1}^{n-1}\cos^{ 2}(\theta_{v })\end{array}\eeq where  \( \prod_{i=l}^{m }v_{i} \triangleq1, \rm{if} ~l>m.\) Assume that \eqref{InductionAssumption} is true for \(n=m\in\nat\),
then, for $m+1$ \eqref{InductionAssumption} and \eqref{yvecDefinition} yields
\beq\nonumber\begin{array}{lll}
\sum_{i=q}^{m}\sin^{2}(\theta_{i}) \prod_{v=i+1}^{m}\cos^{2}(\theta_{v}) =\sum_{ i=q}^{m-1} \sin^{2}(\theta_{i}) \prod_{v=i+1 }^{m}\cos^{ 2}(\theta_{v})+\sin^{ 2}(\theta_{m}) \prod_{v=m+1}^{m}\cos^{2}( \theta_{v})\\~~~~~= \cos^{2}( \theta_{m} )\sum_{i=q}^{m-1} \sin^{2}(\theta_{i}) \prod_{v=i+1}^{m-1} \cos^{2} (\theta_{v })+\sin^{2}(\theta_{m}),
\end{array}
\eeq
According to the supposition \eqref{InductionAssumption} \beq
\begin{array}{lll}
\cos^{2}(\theta_{m})\left( 1-\prod_{i=q}^{ m-1} \cos^{2}(\theta_{i}) \right)+\sin^{2}( \theta_{m})=1-\prod_{i=q}^{m}\cos^{ 2}(\theta_{i}),
\end{array}
\eeq
\end{proposition}
 which establishes the desired result. \hfill $\Box$

  By substituting  Proposition
  \ref{proposition8} into \eqref{before proposition8}
 one obtains\beq\label{after proposition8}
\begin{array}{lll}
\vert \tilde{g}_{q,1}(n_{t}-q-1) \vert^{2} \leq  \underbrace{ \left(\begin{array}{l} \sum_{i=1}^{ n_{t}-q} \vert g_{q,i+q}(1)\vert^{2}  \end{array} \right) }_{=Z(q,c_{1})}  \left(1-\prod_{i= c_{0+}q}^{n_{t}-1}\cos^{2}(\theta_{i}) \right)+\vert \epsilon\vert^{2} \underbrace{ \prod _{v=q}^{n_{t}-1} \cos^{2} \left(\theta _v\right)}_{\leq 1}
\end{array}\eeq
thus,
\beq\label{29_5_11_1}\begin{array}{l}
\vert \tilde{g}_{q,1}(n_{t}-q-1) \vert^{2}\leq \left(1-\prod_{i=c_{0}+q}^{n_{t}-1} \cos^{2}( \theta_{i})\right)Z(q,c_{1})+ \vert\epsilon \vert^{2}\end{array}
\eeq
and by summing both sides of \eqref{29_5_11_1} over $q=2,...,n_{t}$  \beqna \label{First Z}\begin{array}{l}
Z(1,c_{1})\leq\sum_{q=2}^{n_{t}}\left(1-\prod_{ i=c_{0}+q}^{n_{t}-1} \cos^{2}(\theta_{i}) \right)Z(q,c_{1})+(n_{t}-1)\vert\epsilon \vert^{2}\\\leq(1-\prod _{i=c_0+2 }^{c_1} \cos ^2( {\theta }_i))\underbrace{ \sum_{q=2}^{n}Z(q,c_{1})}_ {W(1,c_{1}) }+(n_{t}-1)\vert\epsilon\vert^{2}\leq(1-\prod _{i=c_0+2}^{c_1} \cos ^2( {\theta }_i))W(0,0)+(n_{t}-1) \vert\epsilon\vert^{2}\end{array}
\eeqna
where the last inequality is due to   \( P_{c_{1}}=W(1,c_{1})+Z(1,c_{1}), \) $W(0,0)=P_{0}$, and because $P_{k}$ is a monotonically decreasing sequence\footnote{Forsythe and  Henrici \citep{ forsythe1960cyclic} showed that  the sequence $P_{k}$ is a monotonically decreasing sequence.}. It follows that \begin{equation}\label{z approximation}
Z(1,c_{1})=\sin ^2\left(\Psi _{c_{0}+2 ,c_1}\right) W(0,0)+(n_{t}-1)\vert\epsilon \vert^{2}
\end{equation}
where
\beq \label{definition of psi}
\begin{array}{l}\sin ^2\left(\Psi _{c_{l-1}+2,c_l} \right)=1-\prod _{i=c_{l-1}+2}^{c_l} \cos ^2( \tilde{\theta }_i)
\end{array}\eeq
and \(\tilde\theta_{i}\) is an angle that satisfies   $\vert\tilde \theta_{i}\vert \leq\vert \theta_{i}\vert $.
Thus, \begin{equation}
P_{c_{1}}=W(1,c_{1})+Z(1,c_{1}) \leq W(0,0)=P_{0}
\end{equation}
substituting \eqref{z approximation}
we obtain

\begin{equation}
  W(1,c_{1})\leq W(0,0) \cos ^2\left( \Psi _{2,c_1}\right)-(n_{t}-1)\vert \epsilon \vert^{2}
\end{equation}

Now that this relation is  established, it can be applied to \(\Amat_{c_l}\)'s lower \((n_{t}-l)\times (n_{t}-l)\) block-diagonal, thus
\begin{equation} \label{W(l,cl)}
 W(l,c_{l})\leq W(l-1,c_{l-1}) \cos ^2\left(\Psi _{c_{l-1}+2,c_{l}}\right)-(n_{t}-l)\vert \epsilon\vert^{2}
\end{equation}
By substituting \eqref{W(l,cl)} recursively into itself, one obtains
\beq W(l,c_{l})\leq W(0,0) \prod _{j=1}^l \cos ^2\left(\Psi _{c_{j-1}+2,c_j}\right) -\epsilon ^2 \sum _{j=1}^l b_j \prod _{v=j+1}^l \cos ^2\left(\Psi _{c_{v-1}+2, c_v}\right) \eeq
Thus \beq \begin{array}{ll}  Z(l,c_{l})=\sin ^2\left(\Psi _{c_{l-1}+2,c_l}\right) W(l-1,c_{l-1})+(n_{t}-l)\vert \epsilon^{2} \vert \leq W(0,0) \sin ^2\left(\Psi _{c_{l-1}+2,c_l}\right) \prod _{j=1 }^{l-1} \cos ^2\left(\Psi _{c_{j-1 }+2,c_j}\right)\\~~~~~~~~~~~~~~~~~~~~~~~~~~~~~~~~~~~-\vert\epsilon\vert ^2 \sum _{j=1}^{l-1} b_j \prod _{v=j+1 }^{l-1} \cos ^2\left(\Psi _{c_{v-1}+2,c_v }\right)+(n_{t}-l)\vert \epsilon^{2}\vert
\end{array}\eeq
After a complete sweep
\beq\label{Pcn-1}\begin{array}{lll}
P^{2}_{c_{n_{t}-1}}=&\sum _{l=1}^{n_{t}-2} Z(l,c_{ n_{t}-1})+\vert\epsilon\vert^{2} =\sum _{l=1}^{n_{t}-2} Z(l,c_{l})\leq W(0,0)\sum_{l=1}^{n_{t}-2} \sin ^2\left(\Psi _{c_{l-1}+2,c_l}\right) \prod _{j=1}^{l-1} \cos ^2\left(\Psi _{c_{j-1}+2,c_j}\right) \\&-\sum_{l=1}^{n_{t}-2} \vert\epsilon\vert ^2 \sum _{j=1}^{l-1} b_j \prod _{v=j+1}^{l-1} \cos ^2\left(\Psi _{c_{v-1}+2,c_v}\right)
+\vert \epsilon^{2} \vert\sum_{l=1}^{n_{t}-1 }(n_{t}-l)
\end{array}\eeq
where the first equality is due to the fact that for \(k=c_{l}+1,...,c_{n_t}\),  the sum of squares of the $l$th column remains unchanged; thus, \(Z(l,k)=Z(l,c_l), \forall k>c_{l}\).  Similar to proposition \ref{proposition8}, it can be shown that \( \sum _{l=1}^n \sin ^2\left(\tau _l\right) \prod _{j=1}^{l-1} \cos ^2\left(\tau _j\right)=1-\prod _{j=1}^n \cos ^2 \left( \tau _j\right) \). Thus
\begin{equation}\begin{array}{lll} P^{2}_{c_{n-1}}\leq& W(0,0)\left( 1-\prod_{ j=1 }^{n-2}\cos ^2\left(\Psi _{c_{j-1}+2,c_j }\right)\right)-\sum_{l=1}^{n-2} \epsilon ^2 \sum _{j=1}^{l-1} b_j \prod _{v=j+1}^{l-1} \cos ^2\left(\Psi _{c_{v-1}+2,c_v}\right)
+\vert \epsilon^{2} \vert\sum_{l=1}^{ n-1} b_{l} \end{array}\end{equation}
From \eqref{definition of psi} we have
\(\cos ^2\left(\Psi _{c_{l-1}+2,c_l}\right)\geq\prod _{v=c_{l-1}+2}^{ck_l} \cos ^2( \theta _{ v})\), therefore \begin{equation}\begin{array}{lll} P^{2}_{c_{ n-1}}\leq& W(0,0)\left( 1-\prod_{ j=1}^{n-2}\prod _{v=c_{j-1}+2}^{c_j} \cos ^2(\theta_{v})\right)\\&-\sum _{l=1}^{n-2} \vert\epsilon\vert ^2 \sum _{j=1}^{l-1} (n-j) \prod _{v=j+1 }^{l-1} \prod _{r=c_{v-1}+2}^{c_v} \cos ^2\left(\theta _r\right)+ \frac{ \vert \epsilon^{2}\vert(n^2 - n)}{2}  \end{array}\end{equation}
Recall that $\vert\theta_{i}\vert<\pi/4$, therefore
 \begin{equation}\label{EpsolonInequility} \begin{array}{lll}P^{2}_{c_{n-1}}&\leq&  W(0,0)\left( 1-2^{-(n-2) (n-1)/2 } \right) -\vert\epsilon ^2\vert \left(\sum _{l=1}^{n-2}  \sum _{j=1 }^{l-1} (n-j) 2^{\frac{l^2}{2}-l n+\frac{l}{2}+9 n-45}-\frac{ (n^2  -n)}{2} \right) \\ &\leq & W(0)\left( 1-2^{-(n-2) (n-1)/2 } \right)+\vert\epsilon ^2\vert\frac{ (n^2-n)}{2} \end{array} \end{equation}

It remains to relate \(\epsilon\) to the accuracy of the line search  \(\eta\).   Note that the error  \(\epsilon\) in \eqref{EpsolonInequility} is due to   \eqref{OtimalThetaAndPhi} which is a result of the two  finite-accuracy (of \(\eta\) accuracy) line-searches in \eqref{ThetaOptimization}, and \eqref{EstThetaFromPhi}. If \(\eta\) were zero,   \(\Amat_{k}\)'s \(l,m\)  off diagonal entry would   be  zero  after the \(k\)th sweep, i.e. \beq u(\theta_{k}^{\rm J},  \phi_{k}^{\rm J} )=0\eeq
 where
\beq \begin{array}{l}
u(\theta,\phi)\define\vert[\Rmat_{l,m} (\theta,\phi) \Amat_{k}\Rmat_{l,m}^* (\theta,\phi)] _{l,m}\vert^{2}=u_1(\theta,\phi) +u_2(\theta,\phi),\end{array}\eeq  \beq\begin{array}{l}u_1(\theta,\phi)=4 (a_{l,m}^k)^{2} \sin ^2\left(\gamma _{l,m}+\phi \right)\\u_2(\theta,\phi)=\left(2 \cos (2 \theta ) a^{k}_{l,m} \cos \left(\angle a^{k} _{l,m}+\phi  \right) +\sin (2 \theta ) \left(a^{k}_{l, l}- a^{k}_{ m,m}\right)\right)^2
\end{array}
\eeq
and
\((\theta_{k}^{\rm J},\phi_{k}^{\rm J})\) is the  value given in Theorem
\ref{Theorem:TweDimensionalSearchToOneDemenssionalSearch} when substituting $\Gmat=\Amat_k$.
Recall that  \((\hat\theta_{k}^{\rm J},\hat\phi_{k}^{\rm J})\) (see  \eqref{OtimalThetaAndPhi}) is the non optimal value that is obtained  by the two line searches, then \beq\label{abs(epsilon)}\vert \epsilon \vert^{2} =\max _{k} u (\hat\theta^{\rm J}_{k},\hat\phi_{k}^{\rm J})\eeq   The error \(u(\hat\theta _{k}^{\rm J},\hat\phi_k^{\rm J})\) can be bounded because \(\phi_{k}^{\rm J }=\angle a^{k} _{l,m}\), thus \(\hat\phi_{k}^{\rm J}=-\angle a^{k} _{l,m}+ \eta_{\phi}\) where \(\vert\eta_{\phi}\vert< \eta,\) and
\beq\label{FirstdefineU1}
u_{1}(\hat\theta_{k}^{\rm J},\hat\phi_{k}^{\rm J})=4 (a_{l,m}^k)^{2} \sin ^2\left(\angle a^{k} _{l,m}+\hat\phi_{k} ^{\rm J}\right)\leq 4 (a_{l,m}^k)^{2}\eta^{2}\leq 2\Vert\Gmat\Vert\eta^{2}
\eeq
\beq \label{SecondefineU2}u_{2}(\hat\theta_{k}^{\rm J},\hat\phi_{k}^{\rm J})=
\left(2  a^{k}_{l,m} \cos \left(\eta_{\phi}  \right)\cos^{2}(2\hat\theta_{k}^{\rm J}) +\sin(2\hat \theta_{k} ^{\rm J}) \left(a^{k}_{l, l}- a^{k}_{ m, m}\right)\right)^2
\eeq
To bound $u_2(\hat \theta_{k}^{\rm J},\hat \phi_{k}^{\rm J})$,  note  that if \(a_{ll}^{k}=a_{mm}^{k}\), then \(\hat\theta_{k}^{\rm J}=\theta_{k}^{{\rm J} }\in\{0,\pi/4\}\)  since the line search will not miss these points. Now for the case where \(a_{ll}^{k}\neq a_{mm}^{k}\) we have    $\hat\theta_{k}^{\rm J}=\theta_{k}^{s}+ \eta_{ \theta}$ where
  \beq \label{ThetaKS}\theta_{k}^{s}=\frac{1}{2} \tan ^{-1 }\left( x_{k}\right) \eeq
and \beq\label{DefineX} x_{k}=\frac{2 \vert a^{k}_{l,m}\vert \cos (\eta_{\phi} )}{ a^{k}_{m,m}-a^{k}_{ l,l}}\eeq
Note that
\beq\nonumber \begin{array}{llll}u_{2}(\hat\theta_{k}^{\rm J},\hat\phi_{k}^{\rm J})=\left(2 \cos \left(\eta _{\phi }\right) a_{l,m}^k \left(\cos \left(2 \theta _k^s\right)-2 \eta _{\theta } \sin \left(2 \theta ^*\right)\right)
+\left(a_{l,l}^k-a_{m,m}^k \right) \left(\sin \left(2 \theta _k^s\right)+2 \cos \left(2 \theta ^*\right) \eta _{ \theta} \right)\right)^2
\end{array}\eeq
where \((\theta^{*},\phi^{*})\) is a point on the line that   connects   the points  \((\theta_{k}^{ \rm J},\phi^{\rm J}_{k})\),  \((\hat\theta_{k}^{\rm J},\hat\phi_{k}^{\rm J}) \). By substituting \eqref{ThetaKS} we obtain
\beq \begin{array}{l} u_{2}(\hat\theta_{k}^{\rm J},\hat\phi_{k}^{\rm J})=\left(\frac{2 \cos \left({\eta }_{\phi }\right) a_{l,m}^k+x_k a_{l,l}^k-x_k a_{m,m}^k}{\sqrt{x_k^2+1}}- 4 \eta _{\theta } \sin \left(2 \theta ^*\right) \cos \left({\eta }_{\phi }\right) a_{l,m }^k+2 \eta _{\theta } \cos \left(2 \theta ^*\right) \left(a_{l,l}^k-a_{m,m}^k\right) \right)^2 \end{array}
\eeq
  Using  \eqref{DefineX} and  the fact that the sinusoidal is bounded by  one, and because \(\vert\eta_{\theta}\vert\leq \eta,\) it follows that
\beq\label{BoundU2}\begin{array}{lll}
u_{2}(\hat \theta_{k}^{\rm J},\hat\phi_{k}^{\rm J})\leq4\eta^{2}\left(2   \vert \sin(2\theta^{*})\vert a_{l,m}^k+  \cos(2\theta^{*})\left\vert a_{l,l}^k-a_{m,m}^k \right\vert\right)^2 \\\leq4\eta^{2}\left(4\sin^{2}(2 \theta^{*})\vert a_{l,m}^{k}\vert^{2+} 2\sin(4 \theta^{*})\vert a_{l,m}^{k}\vert\vert a_{ll}^{k}-a_{mm}^{k}\vert+\cos^{2}(2 \theta ^{ *})\vert a_{ll}^{k}-a_{mm}^{k}\vert^{2} \right)\\ \leq4\eta^{2}\left(2 \vert a_{l,m}^{k}\vert^{2} +2\sin(4\theta^{*})\vert a_{l,m}^{k}\vert\vert a_{ll }^{k}-a_{ mm}^{k}\vert+\vert a_{ll}^{k}-a_{mm}^{k}\vert^{2}+2\vert a_{l,m}^{k} \vert^{2} \right) \end{array}\eeq \beq \begin{array}{lll} u_2{(\hat\theta_{k}^{\rm J},\hat\phi_{k }^{\rm J})}\leq 4\eta^{2}\left(2\Vert \Gmat\Vert^{2}+  \sqrt 2\Vert \Gmat\Vert\Vert \Gmat\Vert+ \Vert \Gmat\Vert^{2} \right)\end{array}\eeq
Thus
\beq \label{BoundingEpsilonWithEta}
\begin{array}{lll}
\vert \epsilon \vert ^{2} = \max_{k}u(\hat\theta_{k}^{\rm J},\hat\phi_{k}^{\rm J})\leq2(7+2\sqrt 2 )\eta^{ 2} \Vert\Gmat\Vert^{2}
\end{array}\eeq
This expression is substituted into \eqref{EpsolonInequility}
and the desired result follows.
 \hfill $\Box$

\section{ }\label{ProofOfTheoremQaudraticConvergence}
Without loss of generality, we assume that $W(0,0)\leq \delta^{2}/8$ where $W(k,l)$ is defined in \eqref{DefineW(l,k)}\footnote{ I.e. let $k_0$ be the smallest integer such that $P_{k_0}<\delta^{2}/8$ and $(l_{k_0},m_{k_0})=(1,2)$,  we set $\Amat_{k_0}=\Gmat$).}.  We first prove the theorem assuming that  \(\Gmat\)'s eigenvalues are all distinct. From \eqref{tilde g}  it follows that
\beq\label{before proposition8 monified}
\begin{array}{lll}
\vert \tilde{g}_{q,1}(n_{t}-q-1) \vert^{2} \leq\sum _{j=q}^{n_{t}-1} \sin ^2\left(\theta _j\right) \left|g_{q,j+1}(1)\right|{}^2+ \epsilon ^2 \prod _{v=q}^{n_{t}-1} \cos ^2\left(\theta _v\right)
\end{array}\eeq
Similar to the derivation of \eqref{29_5_11_1}, but without applying Proposition \ref{proposition8}, one obtains
\beq\label{8_6_11_1}\begin{array}{lll}
\vert \tilde{g}_{q,1}(n_{t}-q-1) \vert^{2} \leq Z(q,c_{1})\sum _{j=q}^{n_{t}-1} \sin ^2\left(\theta _j\right)+ \vert \epsilon\vert^{2}
\leq Z(q,c_{1})\sum _{j=2}^{n_{t}-1} \sin ^2\left(\theta _j\right)+ \vert \epsilon \vert^{2} \end{array}\eeq
and by summing both sides of \eqref{8_6_11_1} (similar to the derivation of \eqref{First Z}) over $q=2,...,n_{t}$ it follows that  \begin{equation}\nonumber\begin{array}{lll}
Z(1,c_{1})\leq\left(\sum _{j=2}^{n_{t}-1} \sin ^2\left(\theta _j\right) \right) \underbrace{ \sum_{q=2}^{n_{t}}Z(q,c_{1})}_ {W(1,c_{1})}+(n_{t}-1)\vert\epsilon \vert^{2 }\leq\left(\sum _{j=2}^{n_{t}-1} \sin ^2\left(\theta _j\right) \right) W(0,0)+(n_{t}-1)\vert\epsilon\vert^{2}
\end{array}
\end{equation}

Now that we have established this relation we can apply  it  to the  reduced \(n_{t}-l+1\) lower block diagonal matrix and obtain \(Z(l,c_{l})\leq\left(\sum _{j=c_{l-1 }+1}^{c_{l}} \sin ^2\left(\theta _j\right) \right) W(0,0)+(n_{t}-l)\vert \epsilon\vert^{2}
\).
After a complete sweep we have
 \beq\label{CopleteCycle}\begin{array}{lll}
P^{2}_{c_{n_{t}-1}}\leq&\sum _{l=1}^{n_{t}-2} Z(l,c_{n_{t}-1} )+\vert\epsilon\vert^{2} =\sum _{l=1}^{n_{t}-2} Z(l,c_{l})+\vert\epsilon \vert^{2} \\&\leq W(0,0)\sum _{j=1}^{n_{t}(n_{t}-1) /2} \sin ^2(\theta _j)+\vert \epsilon^{2} \vert\sum_{l=1}^{ n_{t}-1}(n _{t}-l)
\end{array}\eeq
We  now  relate \(\sum_{j=1}^{n_{t}(n_{t}-1)/2} \sin ^2(\theta _j)\)  to $W(0,0)$ ( recall that  \(P_{0}^{2}=W(0,0)\)).  Note that $|
 {a_{ll}^k - a_{mm}^k} |^{2} = | a_{ll}^k - \lambda _l - a_{mm}^k + \lambda _m +
\lambda _l - \lambda _m |^{2} \geq | \lambda _l - \lambda _m |^{2} - | a^{k}_{ll} - \lambda _l |^{2} - | a^{k}_{mm} - \lambda _m|^{2}$, furthermore, by
\citep[][Theorem 1]{HenficSpeed1958}, there exists a permutation to \(\{\lambda_{i}\}_{i=1}^{n_{t}}\) such that
\beq\label{HenrichTheorem}
\vert a_{ii}^{k}-\lambda_{i}\vert\leq \sqrt 2 P_{k},
\eeq
thus,
 \beq \label{EqtationDistance}\vert a^{k}_{ii}-\lambda_{i}\vert \leq\delta/2,\eeq and
\beq \label{101}\vert a^{k}_{ll}-a^{k}_{mm }\vert\geq2\delta-\delta/2-\delta/2=\delta.
\eeq
Recall that the optimal rotation   angle satisfies  \(\tan(2\theta_{k}^{ {\rm J}})= 2\vert a^{k}_{l_{k}m_{k}} \vert/\vert a^{k }_{l_{k}l_{k}}- a_{m_{k }m_{k}}\vert\) while the actual  the rotation angel is   \beq\label{ActualRotation}\hat\theta^{\rm J}_{ k}=\theta_{k }^{{\rm J}}+\eta_{\theta}\eeq It follows that     \beq \label{BoundSinTetak}
\begin{array}{lll}\vert\sin ^{2}(\hat\theta^{\rm J}_{ k}) \vert\leq \vert\sin^{2}( \theta_{k}^{\text{J }})\vert +\vert \eta_{\theta} \sin(2\theta_{ k}^{{\rm J}})\vert \leq\frac{1}{4}\vert2 \theta _k\vert^{2}
+\vert \eta_{\theta}\vert \tan(2\theta_{k }^{{\rm J} } ) \vert\leq\frac{1}{2^{2}} \tan^{2}(2\theta_{k}^{\text{J  }})+\vert \eta_{\theta}\vert \tan(2\theta_{ k}^{ {\rm J} } ) \vert
\\~~~~~~~~~~~\leq\frac{\vert a^{k}_{l_{k},m_{k}} \vert^{2 }}{\delta^{2}}+2\vert\eta_{\theta}\vert \frac{\vert a^{k}_{l_{k},m_{k}}\vert}{\delta}\leq \frac{\vert a^{k}_{l_{k},m_{k}} \vert^{2 }}{\delta^{2}}+ \frac{2\vert\eta_{\theta}\vert\sqrt{ W(0,k)}\vert}{\delta} \end{array}\eeq
Therefore \beq \label{BoundSumSin}\begin{array}{lll}\sum _{k=1}^{n_t(n_t-1)/2} \sin ^2(\hat\theta _k^{\rm J})\leq&\sum _{k=1}^{n_t(n_t-1)/2} \left( \frac{\vert a^{k}_{l_{k},m_{k}} \vert^{2 }}{\delta^{2}}+ \frac{2\vert\eta_{\theta}\vert\sqrt{ W(0,k)}\vert}{\delta} \right)=\frac{1}{\delta^{2}} W(0,k)+\frac{ \eta_{\theta}  (n_{t}^{2}-n_{t})}\delta \sqrt{W(0,k)}
 \end{array} \eeq
By substituting \eqref{BoundSumSin} into \eqref{CopleteCycle} one  obtains
\beq\label{PcBound}\begin{array}{l}
P^{2}_{c_{n_{t}-1}}\leq W(0,0)\left(\frac{1}{ \delta^{2}}W(0,0)+ \frac { (n_{t}^{2}-n_{t})\vert \eta_{\theta}\vert}\delta \sqrt {W(0,k)}\right)+\frac{\vert \epsilon^{2} \vert}{2} \left(n_{t}^2-n_{t}\right),
\end{array}\eeq

 It remains to relate \(\eta_{\theta}\) in \eqref{ActualRotation} to  \(\eta\). Recall that the calculation of $\hat \theta_{k}^{\rm J}$ relies on the calculation of \(\hat \phi_{k}^{\rm J}\). Thus,   \(\eta_{\theta}\) depends  on \(\eta_\phi\) as well, as we now show. Form the  proof of \citep[Theorem 2]{Noam2012Blind}  we know that  if an accurate line search were   invoked, it  would  produce  \(\hat\phi^{\rm J}_{k}=-\angle a^{k}_{l,m}\). However, the actual line search yields \(\hat\phi^{\rm J}_{k}=-\angle a_{lm}^{k} +\eta_{\phi}\), where \(\vert \eta_{\phi}\vert\leq \eta\). Thus, \(\theta_{k}\) is obtained by searching the minimum of a  perturbed    version of   \(S(\Amat_{k},\rvec_{l,m}(\theta,\phi_{k}^{\rm J}))\),     i.e. \beq\label{wOfTheta}
\tilde S(\Amat_{k},\rvec_{l,m}(\theta,\hat\phi^{\rm J}_{k}))=h_{k}(\cos ^2(\theta ) a^{k}_{l,l}-\cos (\eta_{\phi} ) \sin (2 \theta ) a^{k}_{l,m}+\sin ^2(\theta ) a^{k}_{m,m})
\eeq
We first assume that \(a_{ll}^{k}\neq a_{mm}^{k}\).  From the  proof of \citep[Theorem 2]{Noam2012Blind},  the optimal value of \(\theta\) is  \(
\theta^{{\rm J}}_{k}=\frac{1}{2} \tan ^{-1}\left(p_{k}\right
),\) where \(x_{k}=\frac{2 \vert a^{k}_{l,m}\vert}{ a^{k}_{m,m}-a^{k}_{l,l}}\). If one takes into consideration the non-optimality of the  line-search  which obtains \(\hat\phi_{k}^{\rm J}\) and ignores the non-optimality of the line search that obtains \(\hat\theta_{k}^{\rm J}\), then the minimizer  of    \eqref{wOfTheta}  would be  \(
\theta^{s}_{k}=\frac{1}{2} \tan ^{-1}\left( x_{k}\cos (\eta_{\phi} )\right)\)
and
\beq
\begin{array}{lll}
\vert\theta^{\rm J}_{k}-\theta_{k}^{s} \vert=\left\vert\frac{1}{2} \tan ^{-1}\left( x_{k}\cos (\eta_{\phi} )\right)-\frac{1}{2} \tan ^{-1}\left( x_{k}\right)\right\vert \leq\frac{\vert\eta_{\phi}  \sin \left(\eta_{\phi} ^*\right) p_k\vert}{\cos ^2\left(\eta_{\phi} ^*\right) p_k^2+1} \leq \eta_{\phi}^{2}\frac{\vert x_{k}\vert}{\cos ^{2}(\eta_{\phi})x_{k} ^{2}+1}
\end{array}\eeq
where \(\vert\eta_{\phi}^{*}\vert\leq\eta_\phi\). It can be shown that \( \frac{\vert x_{k}\vert }{\cos^{2}(\eta_{\phi})x_{k} ^{2}+1}\leq\frac{1 }{\vert \cos(\eta_{\phi})\vert},
\)
and because \(\hat\theta_{k}^{\rm J}=\theta^{\rm J}_{k}+ \theta_{k}^{s}-\theta^{\rm J}_{k}+\eta_{\phi}\) and \(\vert \eta_{\phi}\vert<\eta\),  the accumulated effect of  the finite accuracy of  both line searches  is bounded by
 \(
\eta_{\theta}\leq \eta+ \frac{\eta^{2}}{\vert \cos(\eta)\vert} \).
For sufficiently small  \(\eta\)   (e.g. \(\eta\leq \pi/20\)) we obtain
 \beq \label{BoundingEtaTheta}
 \eta_{\theta}\leq6\eta/5 \eeq
By substituting \eqref{BoundingEtaTheta} and \eqref{BoundingEpsilonWithEta}  into  \eqref{PcBound}    it follows that

\beq\label{104}\begin{array}{lll}
P^{2}_{c_{n_t-1}}\leq W(0,0)\left(\frac{1}{ \delta^{2}}W(0,0)+\eta \frac {6  (n_{t}^{2}-n_{t})}{5\delta} \sqrt {W(0,0)}\right)+(10+2\sqrt 2 )(n_t^{2}-n_t)
\eta^{2}\Vert\Gmat\Vert^{2}
\end{array}\eeq
Thus, as long as \(\eta\) is negligible with respect to  \(W(0,0)\), the BNSLA will have a quadratic convergence rate  for    \(\Gmat \)  whose  all eigenvalues are distinct. This is not sufficient  since we are interested in  a  matrix \(\Gmat\)  \(n_{t}-n_{r}\) with zero eigenvalues.

To extend the proof to the  case where  the matrix \(\Gmat \) has \(n_{t}-n_{r}\) zero eigenvalues and \(n_{r}\) distinct eigenvalues we use the following theorem:
\begin{theorem}[\citep{parlett1998symmetric} \label{TheoremParlett1998Symmetric} Theorem 9.5.1]
Let \(\Amat\) be an \(n_{t}\times n_{t}\) Hermitian matrix  with eigenvalues \(\{\lambda_{l}\}_{l=1}^{n_{t}}\) that satisfy \beq\label{OrderOfLambda}\lambda_{1}\neq \lambda_{2}\neq \cdots \neq \lambda_{n_{r}}\neq
\lambda_{n_{r}+1}=\lambda_{n_{r}+2}= \cdots =\lambda_{n_{t}}=\lambda\eeq Consider the following partition:
\beq\label{PatritionA}
\Amat=\begin{bmatrix}\Amat_1 & \Bmat  \\
\Bmat & \Amat_2 \\
\end{bmatrix}
\eeq  where \(\Amat_{1}\) is  \(n_{r} \times n_{r}\) and \(\Amat_{2}\) is \((n_{t}-n_{r})\times (n_{t}-n_{r})\) and let \(\delta'>0\). If \(\Vert(\Amat_{1}- \lambda \Imat )^{-1}\Vert< 1/\delta'\), then \beq \Vert \Amat_{2}-\lambda\Imat \Vert\leq  \Vert \Bmat \Vert^{2}/\delta'
\eeq
\end{theorem}

 To apply Theorem
\ref{TheoremParlett1998Symmetric}  to the modified OBNSLA,  we need  to  show that $\Amat_{k}$  satisfies its conditions. This   however is only satisfied by $\Amat_{k}$ with $k\geq m+1$. To show this, note that  \eqref{EqtationDistance} and \eqref{101} are satisfied by \(\Amat_{k}, k\leq m\) for some permutations of the eigenvalues. Thus, due to  the permutation in \eqref{Eq26}, \(\Amat_{k}, k>m\) satisfies  \eqref{EqtationDistance} and \eqref{101}, for the ordering of \eqref{OrderOfLambda}. For the rest of the proof, it is assumed that  \(k>m\).  Let \(\Amat_{ 1}^{k}, \Amat_{2}^{k},\Bmat^{k}\) be \(\Amat_{k}\)'s  submatrices that correspond to the partition in
\eqref{PatritionA}.
   Recall that in our case, \(\lambda =0\), thus,     \eqref{EqtationDistance}  implies  that
\(\Vert \Amat_{1}^{k} \Vert> \delta,\) and also implies that \( a_{ll}^{k}\geq 5\delta/2, \forall 0\leq l\leq n_{r}\). Furthermore, by \citep[Corollary 6.3.4]{horn_and_johonson}
\(
\vert\lambda_l(\Amat_{1}^{k})- a_{ll}^{k} \vert\leq \Vert \Amat_{k} \Vert_{\rm off}\leq \delta/2
\),
thus
\beq
\lambda_{l}(
\Amat_{1}^{k})>0,~\forall 0\leq l \leq n_{r}
\eeq and therefore, the matrix \(\Amat_{1}^{k}\) is invertible, and because \(\Vert \Amat_{1}^{k} \Vert> \delta,\) it follows that  \(\Vert(\Amat_{1}^{k})^{-1}\Vert \leq 1/\delta\), and by   Theorem \ref{TheoremParlett1998Symmetric} we obtain  \beq \label{BoundA2WithB} \Vert\Amat_{2}^{k}\Vert\leq\Vert B_{k}\Vert^{2}/\delta\eeq

   To show that \eqref{BoundA2WithB} leads to quadratic convergence, one must show that the affiliation  of the diagonal entries in the upper \(n_{r}\times n_{r}\) -block of \(\Amat_{k}\) remains unchanged and that the eigenvalue that correspond are arranged in  decreasing order, i.e.  \beq \label{UnChanged} l=\arg\min_{1\leq m\leq n_t}\vert\lambda_{l}-a^{k}_ {mm}\vert =\arg\min_{1\leq m\leq n_t}\vert \lambda_{l}-a^{k+1 }_{mm} \vert,\;\forall l\in \{1,...,n_{r} \}\eeq
and  \beq \label{OrderEigenValue}\lambda_{l}\geq \lambda _{m},\;\forall l\leq m\eeq To  show \eqref{UnChanged}, note that \begin{equation} \label{113_6_8_2012}
\left\vert  a_{l_{k},l_{k}}^{k}-a_{m_{k},m_{k}}^{k
} \right\vert^{2}\leq\sin ^2\left(\theta _{k}\right) \left(2 \cos (\theta_{k} ) a_{l_{k},m_{k}}^{k} \cos \left(\phi _{k}-\angle{ a_{l_{k},m_{k}}^{k}}\right)+\sin \left(\theta _{k}\right) \left(a_{l_{k}, l_{k}}^{k}-a_{m_{k},m_{k}}^{k}\right) \right)^2
\end{equation}
 and that for every \(\theta_{k}\) such that \(l_{k}\leq n_{r}\),   \eqref{BoundSinTetak} is satisfied. Thus
\beq\label{BigInEquality}\begin{small}\begin{array}{lll}
\left\vert  a_{l_{k},l_{k}}^{k}-a_{ l_{'k},l_{k} }^{k+1} \right\vert^{2}\leq  \sin ^2\left(\theta _{k}\right) \left(a_{l_{k},m_{k} }^{k} +\sin \left(\theta _{k}\right) \left(a_{l_{k},l_{k}}^{k}-a_{m_{k},m_{k}}^{k} \right)\right){}^2\\~~\leq  \sin ^2\left( \theta _{k}\right) \left(a_{l_{k},m_{k}}^{k} + \left(\left(a_{l_{k},m_{k}}^{k}\right)^2/ \delta ^2+2 \eta _{\theta } \left.a_{l_{k},m_{k} }^{k}\right/\delta \right){}^{1/2} \delta \right){}^2\\ ~~\leq  \sin ^2 \left( \theta _{k}\right) \left(a_{l_{k},m_{k}}^{k} + \left(\left(a_{l_{k},m_{k}}^{k}\right){}^2+2 \delta \eta _{\theta } a_{l_{k},m_{k} }^{k}\right){}^{1/2} \right){}^2  \leq \text{  }\sin ^2\left(\theta _{k}\right) \left(a_{l_{k},m_{k}}^{k} + \left(\left. \delta ^2\right/4+2 \delta \eta _{ \theta } \delta /2\right){}^{1/2} \right){}^2 \\ ~~\leq \text{  }\sin ^2\left(\theta _{k}\right) \left( a_{l_{k}, m_{k}}^{k} + \delta \left(1/4+ \eta _{ \theta } \right){}^{1/2} \right)^2  \leq  \frac{\delta ^2}{4}\left( 1+4\eta _{\theta }\right) \left(1/2+ \sqrt{1/2+\eta _{ \theta }} \right) ^2
\end{array}\end{small}
\eeq
By restricting  \(\eta\leq1/100\) and considering \eqref{BoundingEtaTheta} it follows that
\beq
\vert a_{l_{{k}},l_{{k}}}^{{k}}-a_{l_{{k}},l_{{k} }}^{{k}+1}\vert \leq 0.65 \delta
\eeq
which establishes \eqref{UnChanged}.

Now that \eqref{UnChanged} is established, \eqref{OrderEigenValue} immediately follows and   for every      \(l,m\)  such that \(l\neq m\) and \(  1\leq l\leq n_{r}\), we have  \(\vert a^{{k}}_{ll}-a^{{k}}_{mm} \vert\geq \delta\).
And   \eqref{CopleteCycle} can be written as
 \beq
 \label{NotCopleteCycle}
 \begin{array}{lll}
P_{c_{n_{r}}+m}\leq&\sum _{l=1}^{n_{t}-1} Z(l,c_{n_{r}}+m)+\vert \epsilon^{2} \vert\sum_{l=1}^{n_{r}}(n_{t}-l) \\&\leq W(0,m)\sum _{j=1}^{c_{n_{r}}} \sin ^2(\theta _j)+\sum_{l=n_{r}+1 }^{n_{t}-1}Z(l,c_{ n_{r}}+m)+\vert \epsilon^{2} \vert\sum_{l=1}^{{n_{ r}}}(n_{t}-l)
\end{array}\eeq
Recall that \(\vert \epsilon\vert^{2}\leq max_{{k}}u(\theta_{{k}},\phi_{{k}})\) where \(u_{1}(\theta_{{k}},\phi_{{k}})\) and \(u_{2}( \theta_{{k}},\phi_{{k}})\) are defined in
 \eqref{FirstdefineU1} and \eqref{SecondefineU2}.
From \eqref{FirstdefineU1} and \eqref{BoundU2}
\(
u(\theta_{{k}},\phi_{{k}})\leq 4\eta^{2}(5P_{{k}}^{2}+ 4 P_{{k}}\Vert\Gmat\Vert+\Vert \Gmat \Vert^{2})
\).
Because \(\vert a^{{k}}_{l_{k}l_{k}}-a^{{k}}_{m_{k}m_{k}} \vert\geq \delta\) for $m< k \leq m+c_{n_r}$, \eqref{BoundSumSin} is satisfied and   similar to \eqref{104} we obtain
\beq
\label{111}
\begin{array}{lll}
P_{c_{n_r}+m}\leq W(0,m)\left( \frac{1}{ \delta ^{2}} W(0,m)+\eta \frac { (n_{t}^{2}-n_{t})}\delta \sqrt {W(0,{k})} \right)\\ +2\left(2n_tn_{r}-n_r^{2}-n_r \right) \eta^{2}(5W(0,m)+ 4 \sqrt{W(0,m)}\Vert \Gmat \Vert+\Vert \Gmat \Vert^{2})+\sum_{l=n_{r}+1}^{ n_{t}-1}Z(l,c_{n_{r}}+m)
\end{array}
\eeq

It remains to bound the term  \(\sum_{ l=c_{n_{r} +1}}^{n_{t}-1} Z(l,c_{n_{r }}+m)\).   Note that for every \(\theta_{ {k}}\) such that \(m< {k}\leq c_{n_{r}}+m\),   \eqref{BoundSinTetak} is satisfied. Let \(Q=\left\{(l,m):1\leq l\leq n_{r}<m\leq n_{t} \right\}\). Note that for   every \({k}\) such that  \((l_{k},m_{k} )\in Q\), \(a_{l_{k}l_{k}}^{{k}}\) and \(a_{m_{k}m_{k}}^{{k}}\)  are located in \(\Amat_{1}^{{k}}\) and \(\Amat_{2}^{{k}}\) respectively. Thus,     \beq\begin{array}{lll}  \vert a_{q, m_{{k}}}^{{k}+1}\vert^{2} \leq \vert a^{{k}}_{ q,m_{{k}}}\vert^{2} +\sin^{2}( \theta_{{k}})\vert a^{{k}}_{l_{{k}},q} \vert^{2 }, \;{\rm for }\; n_r<q<m_{{k}}\\\vert a_{m_{{k}},q}^{{k}+1}\vert^{2} \leq \vert a^{{k}}_{ m_{{k}},q}\vert^{2} +\sin^{2}( \theta_{{k}})\vert a^{{k}}_{l_{{k}},q} \vert^{2 }, \;{\rm for }\; m_{{k}}<q\leq n_{t}  \end{array} \eeq
and from \eqref{BoundSinTetak} \beq\begin{array}{lll}\vert a_{m_{{k}},q}^{ {k}+1}\vert^{2} \leq \vert a^{{k}}_{m_{{k}},q}\vert^{2} +\left( \frac{\vert a^{{k}}_{l_{{k}},m_{{k}} } \vert^{2 }}{\delta^{2}}+2\eta_{ \theta} \frac{ \vert a^{{k}}_{l_{{k}},m_{{k}}}\vert}{ \delta} \right)\vert a^{{k}}_{l_{{k}},q} \vert^{2 } \;{\rm for }\; m_{{k}}<q\leq n_{t}  \\ \vert a_{q,m_{{k}
}}^{{k}+1}\vert^{2} \leq \vert a^{{k}}_{q,m_{ {k}}}\vert^{2} +\left(\frac{\vert a^{{k}}_{l_{{k}} ,m_{{k}}} \vert^{2 }}{\delta^{2}}+2\eta_{
\theta} \frac{\vert a^{{k}}_{l_{{k}},m_{{k}}}
\vert}{ \delta}\right)\vert a^{{k}}_{l_{{k}},q} \vert^{2 }, \;{\rm for }\; n_{r}<q<m_{{k} }
\end{array}\eeq
These can be bounded by
\beq\label{BoundingakPluse1}\begin{array}{lll}
\vert a_{m_{{k}},q}^{{k}+1}\vert^{2},\vert a_{
q,m_{{k}}}^{{k}+1}\vert^{2}\leq W^{2}(0,m) \left( 1+\frac{1}{\delta^{2} } \right)+ \frac{2\eta_{
\theta}}{\delta}W^{3/2}(0,m)
\end{array}
\eeq

Thus, for every \( k\in \{m,...,m+c_{n_{r}}\}\), \beq\begin{array}{lll}
\sum_{l=n_{r+1}}^{n_{t}-1}Z(l, c_{n_{r}}+m)= \sum \limits_{q=n_{r}+1}^{n_{t}-1} \sum \limits_{t= q+1}^{n_{t}} \vert a^{{k}}_{q,t}\vert^{2}\leq O \left(\left( \frac {W(0,0+m)}{\delta} \right)^{2}\right)+O \left(\left( \frac {\eta_{\theta}  W^{3/2}(0,0+m)}{\delta} \right) \right)
\end{array}\eeq
This, together with \eqref{111} and \eqref{BoundingEtaTheta} show  \beq\label{121_7_4_2012}\begin{array}{lll} P^{2}_{c_{n_r}}\leq  O \left(\left( \frac {W(0,0+m)}{\delta} \right)^{2}\right)+O \left(\left( \frac {\eta  W^{3/2}(0,0+m)}{\delta} \right) \right) +O \left(\left( \frac {\eta^{2}  W^{1/2}(0,0+m)}{\delta} \right) \right)+2\left(n_t^{2}-n_t \right) \eta^{2}\Vert\Gmat\Vert^{2} \end{array}\eeq
Since \(P_{{k}}\) is a decreasing sequence, the desired result follows.  \hfill $\Box$

 \section{}
 \label{Appendix:ProofOfClusters}  We first prove the theorem for the case where the non-clustered eigenvalues are the largest; i.e.,     $\lambda_{i}\geq  \lambda_{i+1}+\delta_{c}$   and $\lambda_{i}-\lambda\geq\delta_{c} $ for $i=1,...,n_{r}-v$. Note that $\lambda_{i}=\lambda+\xi_{i-n_{r}-v}$ for $i\in L_{2}=\{n_{r}-v+1,...,n_{r}\}$ and $\lambda_{i}=0$ for $i=n_{r}+1,...,n_{t}$.  Without loss of generality, we assume that $W(0,0)\leq \delta_{c}^{2}/8$ where $W(k,l)$ is defined in \eqref{DefineW(l,k)}.  Let \(\Vmat_{k} \Lambda \Vmat_{k}^{\star}= \Amat_{k}\) be \(\Amat_{k}\)'s EVD, and let \(
\tilde \Amat^{k}=\Vmat_{k}\tilde \bLambda\Vmat^{\star}_{k}\),
\(\hat \Amat^{k}=\Vmat_{k}\hat\bLambda\Vmat^{ \star}_{k}
\)
where
\beq\begin{array}{lll} \tilde \bLambda={\rm diag}(\lambda_{1},\cdots,\lambda _{n_{r}-v},\underbrace {\lambda\cdots \lambda}_v,\underbrace {0 \cdots 0}_{n_{t}-v-n_{r}} )\\ \hat \bLambda={\rm diag}(\underbrace {0\cdots 0}_{n_{r}-v},\xi_{1},\cdots, \xi_{v},\underbrace {0\cdots 0}_{n_{t}-n_{r}-v}, )\end{array}  \eeq
Let  $L_1=\{1,...,n_r-v\}$, $L_3=L\setminus(L_1\cup L_2)$ and  $L_s=(L_1\times L)\cup (L_2\times L_3)$, \(L_c=L_s \cap\{(l,m):l<m\}\).    By combining \eqref{HenrichTheorem} and the condition \(P_{k}^{2}<\delta_{c}^{2}/8\),  it follows that  \eqref{EqtationDistance} and  \eqref{101} hold for \(\delta=\delta_{c}\). Thus, due to the permutation in \eqref{Eq26}, the inequalities  \eqref{EqtationDistance} and  \eqref{101} are satisfied for $\Amat_{k}, k>m$,   $ \forall(l,m)\in L_c$ and $\delta=\delta_{c}$.  In the rest of the proof, we assume that $k>m$. Because $\vert a^{k}_{ll}-a^{k}_{mm}\vert<\delta_{c}, \forall (l,m)\in L_{c}$,  \(\Amat_{k}\) can be partitioned to blocks \(\{\Amat_{ij}^{k}\}_{i,j=1}^{3}\),
where \(\Amat^{k}_{22}\in \comp ^{(n_{r}-v) \times (n_{r}-v)}\), \(\Amat^{k}_{33} \in \comp^{v\times v}\), and  \(\Amat_{ij}\) and \(\Amat_{qt}\) has the same number of rows if \(i=q\) and same number columns if \(j=t\).
In this partition, the   diagonal entries of  $\Amat_{11}^{k}$ are separated      by more than $\delta_c$, and in addition, two   diagonal entries where   each belongs to a different diagonal block (i.e. $\Amat_{11},\Amat_{22},\Amat_{33}$) are also separated by more than  $\delta_c$. Now it is possible to apply \citep[][Lemma 2.3]{VjeranSharp1991} which
asserts that \beq\label{119_30_3}
\Vert \Amat_{ll}^{k}\Vert_{\rm  off}\leq \frac{P_{k}^{2}}{2\delta_{c}}, \text{for } l=2,3,
\eeq
where \(\Vert\cdot\Vert_{\rm off}\) is the sum of squares of the off-diagonal entries.

To show that \eqref{119_30_3} establishes \eqref{quadraticReduction} we first show that the affiliation  of the diagonal entries in the upper $\Amat_{11}^{k}$-block remains unchanged and that no diagonal entry leaves the $\Amat_{22}^{k}$ and $\Amat_{33}^{k}$ blocks; i.e., for $ i=1,2,3$ \beqna \label{permutationCond1} R_{k+1}(v)\in L_i \text{ if } v\in L_i,\text{and}~R_{k}(v)=v \text{ if } v\in L_1,\\ \text{s.t. }  \label{UnChanged2} R_k(v)=\arg\min_{l\in L}\vert\lambda_{v} -a^{k}_{ll}\vert~~~~~~~~~~~~~~~~~~~~~~~  \eeqna
This follows from \eqref{113_6_8_2012}
 and because  for every \(k\) such that  $(l_k,m_k)\in L_c$,    \eqref{BoundSinTetak} is satisfied with replacing $\delta$ by $\delta_c$. Thus, similar to \eqref{BigInEquality},  for every $k$ such that $(l_{k},m_{k})\in L_s$
\(
\left\vert  a_{l_{k},l_{k}}^{k}-a_{ l_{k},l_{k} }^{k+1} \right\vert^{2}\leq  \frac{\delta_c ^2}{4}\left( 1+4\eta _{\theta }\right) \left(1/2+ \sqrt{1/2+\eta _{ \theta }} \right) ^2
\).
By taking  \(\eta\leq1/100\) and considering \eqref{BoundingEtaTheta} it follows that
\(
\vert a_{l_{k},l_{k}}^{k}-a_{l_{k},l_{k }}^{k+1}\vert \leq 0.65 \delta_c,
\)
which establishes \eqref{UnChanged2}; and therefore,  for every \((l,m)\in L_s\),  \(\vert a^{k}_{ll}-a^{k}_{mm} \vert\geq \delta_c\).

Similar  to the derivation of \eqref{111},
\beq
\label{111_2}
\begin{array}{lll}
P_{c_{n_t-v-r}+m}\leq W(0,m)\left( \frac{1}{ \delta ^{2}} W(0,m)+\eta \frac { {n_{t}^{2}-n_{t}}}\delta_{c} \sqrt {W(0,k)} \right)\\ ~+4c_{n_{t}-v-r}  \eta^{2}(5W(0,m)+ 4 \sqrt{W(0,m)}\Vert \Gmat \Vert+\Vert \Gmat \Vert^{2})+
\sum_{l=n_{t}+v+r+1}^{ n_{t}-1}Z(l,c_{n_{t}-v-r}+m)
\end{array}
\eeq
and similar to the derivation of
\eqref{121_7_4_2012}, we obtain  \beq\label{121_7_4_2012_2}\nonumber\begin{array}{l} P^{2}_{c_{n_t-v-r}+m}\leq  O \left(\left( \frac {W(0,m)}{\delta_{c}} \right)^{2}\right)+O \left(\left( \frac {\eta  W^{3/2}(0,m)}{\delta_{c}} \right) \right) +O \left(\left( \frac {\eta^{2}  W^{1/2}(0,m)}{\delta_{c}} \right) \right)+2\left(n_t^{2}-n_t \right) \eta^{2}\Vert\Gmat\Vert^{2} \end{array}\eeq
Since \(P_{k}\) is a decreasing sequence, the desired result follows.
\section{}
\label{Appendix:CorollaryLimSup}
The corollary follows from  Theorem \ref{Proposition Linear Convergence}  and from the following proposition:
\begin{proposition}
Let \(b>0, 0<\rho<1\) and let \(a_{n}\) be a non-negative  sequence that satisfies
\(
a_{n+1}\leq \rho a_{n}+b,~\forall n\in\nat,
\) then,\( \label{limsup}
\lim \sup_{n} a_{n}\leq \frac{b}{1-\rho}
\).
\end{proposition}
\IEEEproof
 We first assume that for some \(n\in\nat \) \(a_{n}\geq \frac{b}{1-\rho}\). In this case we have  \(a_{n+1}\leq a_{n} \) which means that \(a_{n}\)   is a  monotonic decreasing sequence as long as \(a_{n}\geq \frac{b}{1-\rho}\). In the case where \(a_{n}< \frac{b}{1-\rho}\) we have \(a_{n+1}<\frac{b}{1-\rho}\). These mean that  either  \(a_{n}\) converges to a limit \(\xi>\frac{b}{1-\rho}\), or that  it satisfies \(\lim \sup_{n} a_{n}\leq \frac{b}{1-\rho}\). Assume that the previous statement is true, then for every \(\epsilon>0\), there exists \(n_{\epsilon}\in\nat\) such that \( \label{absolute inequality}\xi-\epsilon\leq a_{n}\leq \xi+\epsilon,~\forall n>n_{\epsilon}\).
By substituting it into \(a_{n+1}\leq \rho a_{n}+b\), i.e., substituting \(\xi-\epsilon\)  for \(a_{n+1}\) and \(\xi+\epsilon\) for \(a_{n}\) it follows that for every \(\epsilon>0 \),   \( \xi(1-\rho)\leq b+\epsilon(1+\rho) \). This is equivalent to \(
\xi \leq \frac{b}{1-\rho}+\frac{\epsilon(1+\rho)}{(1-\rho)},~ \forall \epsilon>0 \)
which is a contradiction. \hfill $\Box$

\bibliographystyle{ieeetr}

\end{document}